\documentclass[
    twocolumn,
    superscriptaddress,
    longbibliography,
    floatfix,
    showkeys,
    xcolor
    ]{revtex4-1}

\usepackage[colorlinks, linkcolor = black, citecolor = black, filecolor = black, urlcolor = blue]{hyperref}
\usepackage{tabstackengine}

\usepackage{amsmath}
\usepackage{amssymb}
\usepackage{graphicx,braket,nicefrac}
\usepackage{siunitx,upgreek}
\usepackage{mathtools}
\usepackage{color}
\usepackage{hhline}
\usepackage{tabularx}
\usepackage{ulem}
\newcommand{\Ca}{{}^{40}\textrm{Ca}^+}

\newcommand{\level}[2]{{}^{2}\textrm{#1}_{#2}}

\let\oldsqrt\sqrt
\def\sqrt{\mathpalette\DHLhksqrt}
\def\DHLhksqrt#1#2{%
\setbox0=\hbox{$#1\oldsqrt{#2\,}$}\dimen0=\ht0
\advance\dimen0-0.2\ht0
\setbox2=\hbox{\vrule height\ht0 depth -\dimen0}%
{\box0\lower0.4pt\box2}}

\newcommand{\sx}[1]{\hat{\sigma}^{#1}}
\newcommand{\sxj}[2]{\hat{\sigma}^{#1}_{#2}}
\newcommand{\Rx}[1]{\hat{R}^{#1}(\theta)}

\newcommand{\RxT}[2]{\hat{R}^{#1}(#2)}

\newcommand{\Idj}[1]{\hat{\mathbb{I}}_{#1}}

\newcommand{\Sj}[1]{\hat{\textbf{S}}_{#1}}

\newcommand{\SjT}[1]{\tilde{\textbf{S}}_{#1}}
\newcommand{\SxT}[1]{\tilde{S}^{#1}}
\newcommand{\SjxT}[2]{\tilde{S}_{#1}^{#2}}

\usepackage{lipsum}

\begin{document}

\title{Constructing the spin-1 Haldane phase on a qudit quantum processor}

\author{C. L. Edmunds}
\email{edmunds.claire@gmail.com}
\affiliation{Universit\"{a}t Innsbruck, Institut f\"{u}r Experimentalphysik, Technikerstra\ss e 25a, Innsbruck, Austria}

\author{E. Rico}
\email{enrique.rico.ortega@gmail.com}
\affiliation{EHU Quantum Center and Department of Physical Chemistry, University of the Basque Country UPV/EHU, P.O. Box 644, 48080 Bilbao, Spain}
\affiliation{Donostia International Physics Center, 20018 Donostia-San Sebastián, Spain}
\affiliation{IKERBASQUE, Basque Foundation for Science, Plaza Euskadi 5, 48009 Bilbao, Spain}

\author{I. Arrazola}
\affiliation{Instituto de F\'{i}sica Te\'{o}rica, UAM-CSIC, Universidad Aut\'{o}noma de Madrid, Cantoblanco, 28049 Madrid, Spain}

\author{G. K. Brennen}
\affiliation{Centre for Engineered Quantum Systems, School of Mathematical and Physical Sciences, Macquarie University, NSW 2109, Australia}

\author{M. Meth}
\affiliation{Universit\"{a}t Innsbruck, Institut f\"{u}r Experimentalphysik, Technikerstra\ss e 25a, Innsbruck, Austria}

\author{R. Blatt}
\affiliation{Universit\"{a}t Innsbruck, Institut f\"{u}r Experimentalphysik, Technikerstra\ss e 25a, Innsbruck, Austria}
\affiliation{Alpine Quantum Technologies GmbH, 6020 Innsbruck, Austria}
\affiliation{Institut f\"{u}r Quantenoptik und Quanteninformation, \"{O}sterreichische Akademie
der Wissenschaften, Otto-Hittmair-Platz 1, 6020 Innsbruck, Austria}

\author{M. Ringbauer}
\affiliation{Universit\"{a}t Innsbruck, Institut f\"{u}r Experimentalphysik, Technikerstra\ss e 25a, Innsbruck, Austria}

\date{\today}     

\begin{abstract}
Symmetry-protected topological phases have fundamentally changed our understanding of quantum matter. An archetypal example of such a quantum phase of matter is the Haldane phase, containing the spin-1 Heisenberg chain. The intrinsic quantum nature of such phases, however, often makes it challenging to study them using classical means. Here, we use trapped-ion qutrits to natively engineer spin-1 chains within the Haldane phase. Using a scalable, deterministic procedure to prepare the Affleck-Kennedy-Lieb-Tasaki (AKLT) state within the Haldane phase, we study the topological features of this system on a qudit quantum processor. Notably, we verify the long-range string order of the state, despite its short-range correlations, and observe spin fractionalization of the physical spin-1 particles into effective qubits at the chain edges, a defining feature of this system. The native realization of Haldane physics on a qudit quantum processor and the scalable preparation procedures open the door to the efficient exploration of a wide range of systems beyond spin-1/2.
\end{abstract}

\maketitle

Topological phases of matter have emerged as a new paradigm that leverages the synergy between topological concepts and condensed matter physics with potential applications for novel materials~\cite{Kumar:2021, Huang:2017, Jin:2023}, robust quantum information~\cite{Miller:2015}, and metrology~\cite{Bartlett:2018}. A rich tapestry of topologically distinct phases emerges when extending the concept of topological insulators to consider states that are topologically protected when a specific \textit{symmetry} is preserved, such as time reversal, particle-hole conjugation, or spatial translations. These states constitute Symmetry Protected Topological (SPT) phases, which play a crucial role in defining the allowed phases and extend our understanding beyond the well-established symmetry-preserving and symmetry-breaking phases of matter~\cite{Pollman:2012_feb, Pollman:2012_sep, Pollman:2017}. Typically, states within an SPT phase have high degrees of quantum complexity and are often found in high-dimensional spin spaces, making them challenging to classically simulate~\cite{Ellison:2021}. 
Consequently, we look to physical implementations to simulate these topological materials in an experimentally controlled context~\cite{Lanyon:2011, Meth:2022}. 

A paradigmatic example of an SPT phase was first proposed by F.D.M. Haldane~\cite{Haldane:1983, Haldane:1983_apr, Haldane:2016}, and comprises integer-spin chains. Unlike their half-integer spin counterparts, the spin chains within this Haldane phase are archetypal examples of SPT states, exhibiting interesting properties from both a condensed matter and quantum information perspective. Key examples of states within the Haldane phase include the spin-1 Heisenberg chain and the Affleck-Kennedy-Lieb-Tasaki (AKLT) model~\cite{Affleck:1987} and, curiously, also the spin-1/2 cluster state~\cite{Else:2012}. Research on the entanglement structure of states within the Haldane phase, particularly focusing on the AKLT state, has been an active area of study~\cite{Moudgalya:2018, Chattopadhyay:2020, Pollman:2010, Renard_Ch2:2001}. Expanding on these investigations, quantum experiments using digital or analog simulation have emerged to deepen our grasp of the underlying physics~\cite{Chen:2023, Sompet:2022, Cohen:2014, Cohen:2015, Senko:2015, Wang:2023, Zhou:2021, Lanyon:2011}. Thus far, experimental endeavors to model states within the spin-1 Haldane phase have relied on encoding higher-dimensional physics into spin-1/2 qubit systems, which are then probabilistically projected onto the necessary spin-1 subspace~\cite{Kaltenbaek:2010,Chen:2023, Smith:2023, Sompet:2022, Lake:2009}.

Here we use trapped-ion qudits~\cite{Ringbauer:2022}, which have emerged as an exciting experimental platform~\cite{Kristen:2020, Chi:2022, Ringbauer:2022, Senko:2015} for natively studying high-dimensional spin systems~\cite{Cohen:2014, Cohen:2015, Wang:2023, Zhou:2021}. Using a chain of qutrits in a universal quantum processor, we directly engineer and study spin-1 chains within the Haldane phase. We explore the phase from both a condensed matter and quantum information perspective by realizing two different states from the Haldane phase: the spin-1 AKLT chain created in qutrits and the spin-1/2 cluster state created in qubits, exploiting the flexibility of our trapped-ion processor to encode different dimensional spins. This allows us to observe not only the characteristic short-range correlations and long-range order, but also the fractionalization of fundamental spin-1 particles into effective spin-1/2 degrees of freedom. By directly simulating the SPT states using qudits, we can reduce quantum resource overhead, eliminate any probabilistic post-selection, and remove the complexity of encoding and decoding between $d$-dimensional spins and qubits.

\section{Creating the spin-1 AKLT state in trapped-ion qutrits}

\begin{figure}[t]
    \centering
    \includegraphics[scale=1]{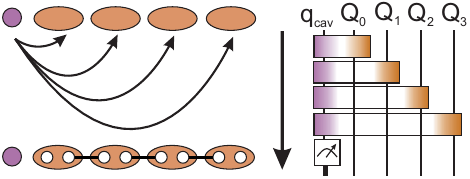}
    \caption{Creation of the AKLT state in trapped-ion qutrits. (Left) The AKLT state is encoded using a chain of spin-1 particles (orange ovals). Each spin-1 can be pictured as two virtual qubits (white circles) projected onto a spin-1 subspace, which, when linked by a singlet on neighboring sites, forms the AKLT state. In our work, there are no physical spin-1/2 degrees of freedom. The singlet bonds are created by sequentially coupling each spin-1 particle to an ancilla qubit (purple circle). 
    (Right) A schematic circuit shows the sequential generation procedure with a linear number of entangling gates between the ancilla qubit, $\textrm{q}_\textrm{cav}$, and the spin-1 qutrits, $\textrm{Q}_j$. A final ancilla measurement projects the qutrits into one of two possible ground states of the AKLT Hamiltonian.}
    \label{fig:fig1}
\end{figure}

We aim to experimentally realize the AKLT state in a native spin-1 system -- a trapped-ion qudit-based quantum processor -- and study the interplay between symmetry and topological properties. The AKLT states are defined as the ground states of a spin-1 Heisenberg Hamiltonian with a quadratic perturbation,
\begin{equation}
\begin{aligned}
\label{eq:ham_AKLT}
    \hat{H}_\textrm{AKLT} 
    &= \sum_{j=1}^{N-1} \left[
    \frac{1}{2} \SjT{j} \cdot \SjT{j+1} + \frac{1}{6}(\SjT{j} \cdot \SjT{j+1})^2 + \frac{1}{3} \right],
\end{aligned}
\end{equation}
where \mbox{$\SjT{j}=\{\SjxT{j}{x}, \SjxT{j}{y}, \SjxT{j}{z}\}$} are the spin-1 matrices acting on the $j^\textrm{th}$ particle. We denote our spin-1 basis as \mbox{$\{\ket{x}, \ket{y}, \ket{z}\}$}; the basis definition and the corresponding spin-1 matrices are found in Appendix B. To experimentally realize the AKLT state, we first note that it can be expressed as a matrix product state (MPS) with bond dimension $D=2$~\cite{Cirac:2021, Schuch:2011}. Here the bond dimension $D$ is a measure of the dimension of the connection between the MPS tensors, representing the sites in the AKLT chain. In Refs.~\cite{Schon:2005,Schon:2007} it was shown that MPS's are equivalent to a class of \textit{sequentially generated states} created by initializing a string of atoms into a product state and sequentially entangling each atom with a shared ancilla system (Fig.~\ref{fig:fig1}). The use of an all-to-all connected ancilla as in the trapped-ion platform ensures that entanglement is spread across the entire chain without the need for physical interaction between every pair of spins. 

We encode spin-1 qutrits in the Zeeman sub-levels of the ground-state $\level{S}{1/2}$ and metastable $\level{D}{5/2}$ energy levels of trapped $\Ca$ ions in a linear Paul trap (see Appendix A for details). The states are encoded as \mbox{$\ket{x}=\level{D}{5/2}\ket{-3/2}$}, \mbox{$\ket{y}=\level{S}{1/2}\ket{-1/2}$}, \mbox{$\ket{z}=\level{D}{5/2}\ket{-1/2}$}. The ancilla is chosen to match the AKLT MPS bond dimension of $D=2$; therefore, it is encoded in a trapped-ion qubit with states $\{\ket{\uparrow}, \ket{\downarrow}\}$. Note that larger bond dimensions could be conveniently realized by choosing the ancilla as a qudit. We now initialize $N$ qutrits and the ancilla qubit into a product state: $\ket{\uparrow, z\dots z}$. The ancilla is subsequently coupled to each of the qutrits, for a total of $N$ two-body interactions. The unitary coupling between the ancilla and each spin-1 particle is derived in Appendix C and can be described by two interactions,
\begin{equation}
    \label{eq:AKLT_seq_unitary}
    \begin{aligned}
        \hat{U}\ket{\uparrow, z}
        &= \ket{\downarrow, x} + i\ket{\downarrow, y} + \ket{\uparrow, z} \\
        \hat{U}\ket{\downarrow, z}
        &= \ket{\uparrow, x} - i\ket{\uparrow, y} -\ket{\downarrow, z}.
    \end{aligned}
\end{equation}
To produce the interaction using the gates available in our experiment, namely two-level resonant rotations and M{\o}lmer-S{\o}rensen entangling gates~\cite{Ringbauer:2022}, we use the circuit synthesis tools in the BQSKit python package~\cite{BQSKit}. The circuit required for each ancilla-qutrit pair can be achieved using two entangling gates and several local rotations, see Tab~\ref{tab:MPS_gate_decomp} in Appendix C.

Once the chain is fully entangled, the ancilla can either be kept in the computation or measured. Projecting the ancilla onto one of two possible states creates one of two degenerate AKLT ground states in the qutrit chain. While we post-select the two outcomes separately in this work, the feed-forward of the ancilla measurement result could be used to deterministically prepare the same AKLT ground state in every shot. In contrast, most qubit-based encodings must project onto a spin-1 subspace, such that the number of accepted measurements for a length-$N$ chain is $(\tfrac{3}{4})^N$~\cite{Chen:2023, Sompet:2022}. The qutrit protocol that we present is highly scalable, requiring only a single ancilla, $2N$ entangling gates, and, when combined with feed-forward conditioned on a single measurement outcome, no post-selection. 

\section{Verifying the topological properties of the spin-1 AKLT chain}

\begin{figure}
    \centering
    \includegraphics[width=86mm]{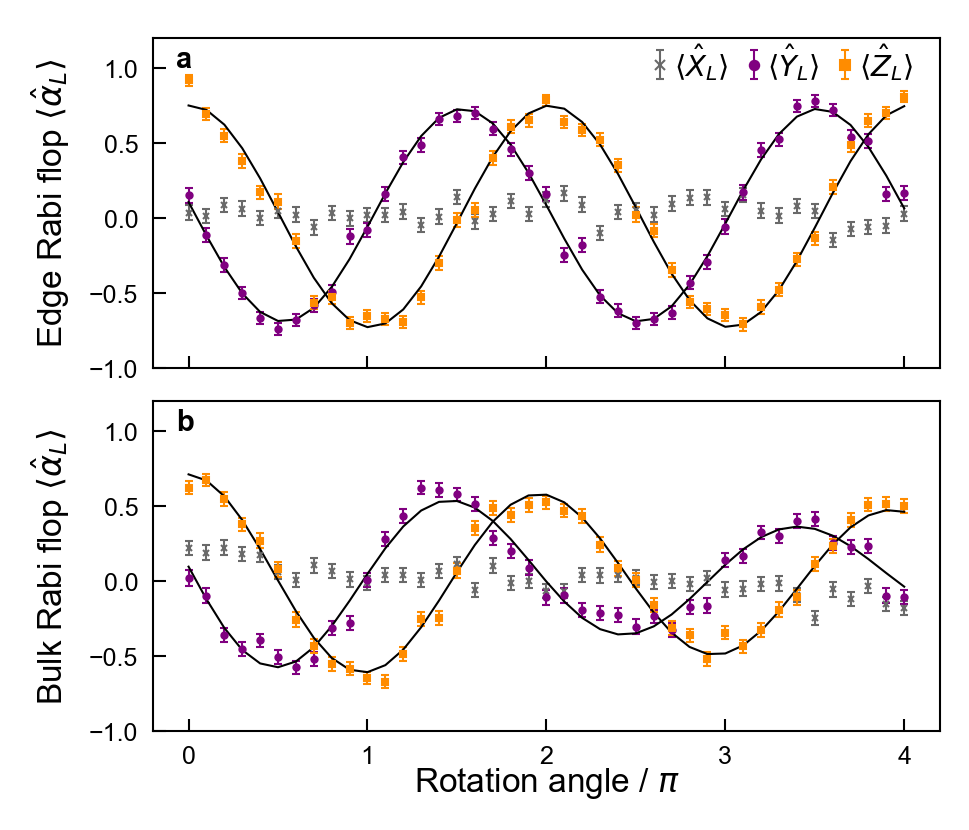}
    \caption{Rabi flops of a fractionalized qubit localized at the edge of an AKLT chain with $N=4$. The rotations are driven using (a) a unitary localized on the edge, generated by the operator $\hat{X}_L$, and (b) a bulk operator driving a global rotation about $\SjxT{j}{x}$ for each spin. 
    The edge operators $\{\hat{X}_L, \hat{Y}_L, \hat{Z}_L\}$ form an SU(2) algebra and are measured for each point. Black lines are sinusoidal fits to the data. (a) For the edge-driven rotations we fit contrasts of 0.71(3) and 0.74(3) for $\langle \hat{Y}_L \rangle$ and $\langle \hat{Z}_L\rangle$, respectively. (b) For the bulk-driven rotations, the corresponding fitted contrasts are 0.69(5) and 0.69(3). In panel (b), an exponential decay $e^{-n\pi/\tau}$ with respect to the rotation angle $n\pi$ gives a decay constant $\tau=14(3)$ and $\tau=30(6)$ for $\langle \hat{Y}_L \rangle$ and $\langle \hat{Z}_L \rangle$, respectively.}
    \label{fig:fig2}
\end{figure}

The topological phase of the AKLT state is robust as long as one of the symmetries of time reversal, inversion, or $\pi$-rotation about a pair of orthogonal axes is preserved. A fascinating consequence of this symmetry protection is quantum number fractionalization, which is expected to emerge when considering a chain with open boundary conditions~\cite{Pollman:2010}. Here, the original spin-1 degrees of freedom fractionalize into two unpaired spin-1/2 degrees of freedom located at the ends of the chain. This effect is described by a projective representation of the $\mathbb{Z}_2\times \mathbb{Z}_2$ symmetry operations at the edge of the chain~\cite{Pollman:2012_feb}. Unlike the unique ground state formed by the Hamiltonian with closed boundaries, these emergent spin-1/2 entities lead to a four-fold degenerate ground-state subspace. Furthermore, spin chains in the Haldane phase possess a finite correlation length, implying the existence of a finite energy gap separating the ground state from excited states (known as the Haldane gap)~\cite{Kennedy:1990}. Despite the short correlation length, states within the Haldane phase exhibit hidden antiferromagnetic order captured by a non-local order parameter~\cite{Perez:2008, Nijs:1989,Pollman:2012_sep}, defined in Eq.~\eqref{eq:string_order} below.

We engineer each of the four degenerate AKLT ground states in our system for chains ranging from $N=2$ to $N=5$ spin-1 sites. To create a particular ground state, we set the initial ancilla state and separate the results based on the measurement of the ancilla qubit. This final step can be performed deterministically using feed-forward on the ancilla measurement. To verify our final state we measure its energy, known to be identically zero for an AKLT ground state. The energy measurement is an effective and scalable verification method for long spin chains, as it requires no knowledge of the ideal state and needs only nine measurements for any length (see Appendix D for details). We measure an energy/site increasing from 0.09(1) for an $N=2$ chain to 0.3(1) for an $N=5$ chain. For short chains, we can compare these results against qutrit state tomography. Utilizing a four-element mutually unbiased basis, we can reconstruct states with up to $N=4$ qutrits, using a scheme analogous to that presented for qubits in Ref.~\cite{Stricker:2022} (see Appendix E for details). We measure fidelities of \mbox{$\{87(3), 76(2), 55(2)\}\%$} for chains of length \mbox{$N = \{2, 3, 4\}$}.

In the AKLT state, edge states with fractionalized spins emerge from excitations localized at the interface or boundary of a material. They are often topologically protected, exhibiting properties distinct from the bulk of the material.  In the AKLT chain, an SU(2) subspace is created when the physically indivisible spin-1 chain fractionalizes to form a lower-dimensional spin-1/2 degree of freedom, which is localized at the edge of the chain. The SU(2) algebra  at one of the edges can be defined by the generators $\hat{X}_L, \hat{Y}_L, \hat{Z}_L$,
\begin{widetext}
\begin{align}
    \label{eq:logic_x}
    \hat{X}_L &= \left[
    (\ket{xx}+\ket{yy}+\ket{zz})(\bra{yz}-\bra{zy}) \right]
    + \frac{i}{2} \left[
    (\ket{xy}-\ket{yx})(\bra{zx}-\bra{xz}) \right] + \textrm{h.c.}\\
    \label{eq:logic_y}
    \hat{Y}_L &= \left[
    (\ket{xx}+\ket{yy}+\ket{zz})(\bra{zx}-\bra{xz}) \right] 
    + \frac{i}{2} \left[
    (\ket{yz}-\ket{zy})(\bra{xy}-\bra{yx}) \right] + \textrm{h.c.}\\
    \label{eq:logic_z}
    \hat{Z}_L &= \left[
    (\ket{xx}+\ket{yy}+\ket{zz})(\bra{xy}-\bra{yx}) \right] 
    + \frac{i}{2} \left[
    (\ket{zx}-\ket{xz})(\bra{yz}-\bra{zy}) \right] + \textrm{h.c.}
\end{align}
\end{widetext}
where \mbox{$[\hat{X}_L, \hat{Y}_L] = 2i\hat{Z}_L$} and h.c. is the Hermitian conjugate. We verify the SU(2) subspace by driving Rabi flops (Bloch-sphere rotations) generated by $\hat{X}_L$, and measure the expectations of the three operators, as shown in Fig.~\ref{fig:fig2}(a) for a length $N=4$ chain. The unitary for each point in the Rabi flop is compiled using the BQSKit circuit synthesis tool~\cite{BQSKit}. As the length of the AKLT chain is increased from two to four sites, the Rabi flop contrast remains approximately constant within \mbox{69\% - 74\%}, despite the fidelity decreasing with string length. This is consistent with the understanding that the spin-1/2 degree of freedom is localized at the edge of the chain, with a rapidly decaying leakage into the bulk. Consequently, the qubit degree of freedom is generated with an approximately constant fidelity, despite the increasing length of the bulk. A complete analysis for chain lengths ranging from $N=2$ to $N=5$ can be found in Appendix F.

We examine the correspondence between the properties of the bulk and the edge physics predicted for any spin chain in the Haldane SPT phase~\cite{Pollman:2010}. There exists a bulk-operator, \mbox{$e^{-i\theta\sum_{j=1}^N\SjxT{j}{x}/2}$}, which is equivalent to the edge-unitary generated by $\hat{X}_L$ when we restrict to the ground-state manifold. The global rotation acts trivially on the bulk, as it contains only singlet states, and only affects the edge degrees of freedom. We show this by applying the bulk operator to the AKLT state and measuring the expectations of the three SU(2) operators, $\hat{X}_L, \hat{Y}_L, \hat{Z}_L$. The resulting Rabi flops, shown in Fig.~\ref{fig:fig2}(b) for an $N=4$ chain, have consistent behavior with those generated by the edge operator. There is a slight decay in the bulk oscillation contrast, as the gate error scales with the rotation angle. In contrast, for the edge operators, a decomposition could be chosen where the gate error is independent of the rotation angle, thus displaying no decay. We note that for both the edge- and bulk-driven rotations, we observe a \textit{continuous} cyclic permutation between the degenerate ground states, highlighting the robustness of the Haldane phase against global rotations of any angle. While the Haldane phase is generally characterized by a $\mathbb{Z}_2\times \mathbb{Z}_2$ symmetry generated by discrete $\pi$-rotations about the three orthogonal axes, for the AKLT ground state the ground subspace is indeed endowed with the full SO(3) rotational symmetry. 

\begin{figure}[h]
    \centering
    \includegraphics[width=86mm]{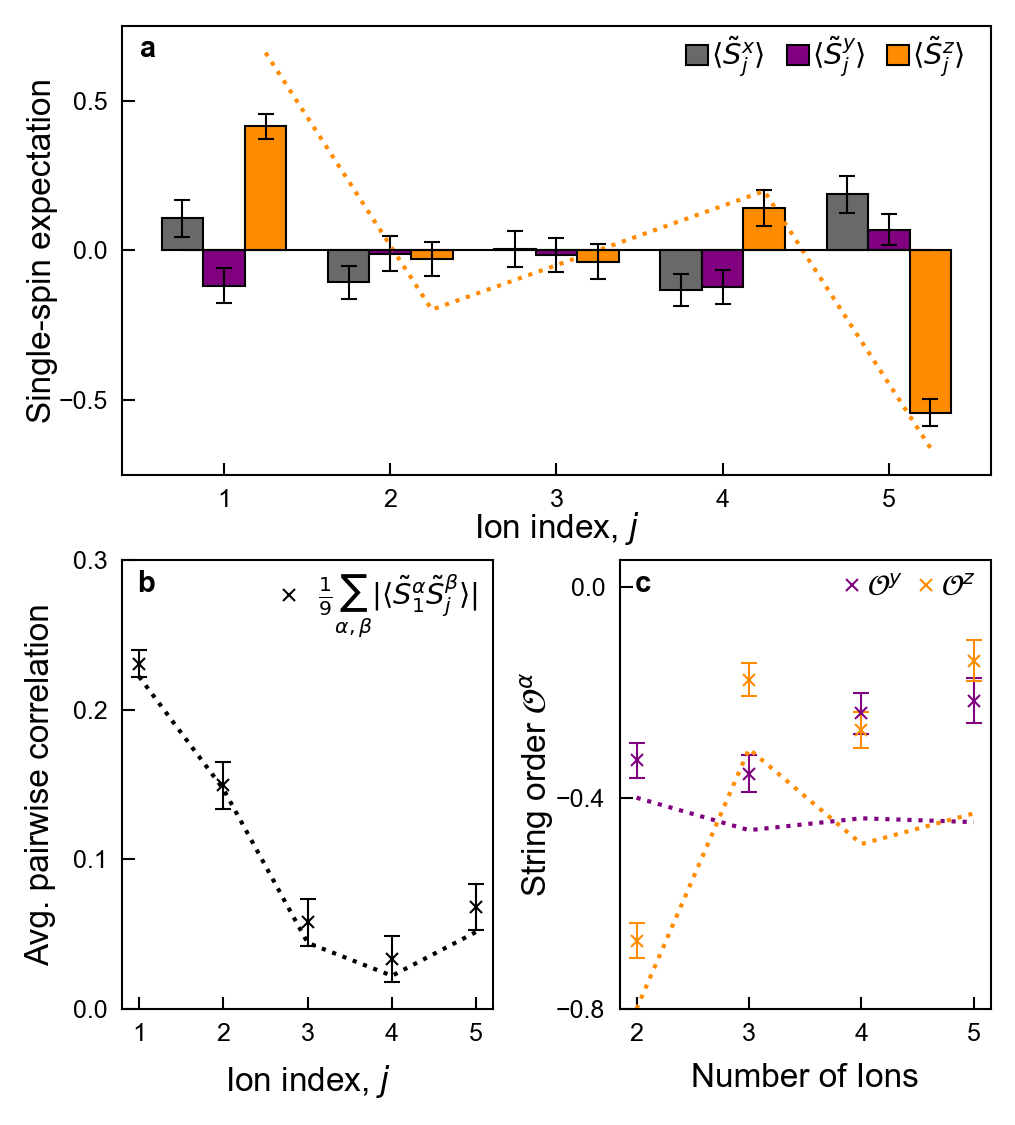}
    \caption{The measured (a) single-qutrit and (b) two-qutrit expectations of the spin-1 operators $\{\SxT{x},\SxT{y},\SxT{z}\}$ are shown for each qutrit in the AKLT state with $N=5$ and final ancilla $\ket{\uparrow}$. In all panels, dashed lines are noise-free circuit simulations, and error bars correspond to one standard deviation from Monte Carlo resampling. In panel (a), the expectation $\langle \SjxT{j}{x}\rangle$ is shown in gray, $\langle \SjxT{j}{y}\rangle$ in purple and $\langle\SjxT{j}{z}\rangle$ in orange. The noise-free simulations for both \mbox{$\langle\SxT{x}\rangle,\langle\SxT{y}\rangle$} are identically zero. In panel (b), the two-spin correlations are measured between the left-most spin with index 1 and a second spin with increasing distance along the chain, $j$. For each ion pair, $\langle\SjxT{1}{\alpha} \SjxT{j}{\beta}\rangle$ is measured for $\alpha,\beta\in\{x, y, z\}$, and the average deviation from zero across the nine measurements is plotted, $\tfrac{1}{9}\sum_{\alpha,\beta = x,y,z} |\langle\SjxT{1}{\alpha} \SjxT{j}{\beta}\rangle|$. (c) The string order parameter defined in Eq.~\eqref{eq:string_order} is measured in two directions for the observables $\mathcal{O}^y$ (purple) and $\mathcal{O}^z$ (orange).
    }
    \label{fig:fig3}
\end{figure}

We can also observe the characteristic properties of the SPT state in terms of its local and non-local order parameters. Figure~\ref{fig:fig3}(a) shows that there is no local order in the bulk of the chain by measuring the expectation of the spin-1 operators $\SjxT{j}{\alpha}$ with $\alpha\in\{x, y, z\}$ for a length $N=5$ chain. The values of $\langle \SjxT{j}{x}\rangle$ and $ \langle \SjxT{j}{y}\rangle$ are observed to be close to zero as expected, with an average deviation of 0.08(3) and 0.03(2), respectively. In contrast, pronounced peaks of $\langle\SjxT{j}{z}\rangle$ at the edges of the chain indicate the existence of edge modes. These modes are located on the outermost qudits with a small leakage to their respective neighbors, due to the finite correlation length of the AKLT chain. Furthermore, we observe the short-range nature of the spin-spin correlations by measuring the two-qutrit expectation values, $\langle\SjxT{1}{\alpha}\SjxT{j}{\beta}\rangle$ for increasing distance in Fig.~\ref{fig:fig3}(b). The results show that the correlation strength decreases rapidly with distance, which indicates a spectral gap above the ground-state manifold~\cite{Hastings:2006}, characteristic of the Haldane phase.

Finally, the hidden antiferromagnetic order of the Haldane phase is revealed by measuring the non-local string order parameter,
\begin{equation}
    \label{eq:string_order}
    \mathcal{O}^\alpha = \langle \SjxT{1}{\alpha} e^{i\pi \prod_{k=2}^{N-1} \SjxT{k}{\alpha}} \SjxT{N}{\alpha} \rangle ,
\end{equation}
for $\alpha = \{y,z\}$, see Fig.~\ref{fig:fig3}(c). The measured values are consistently non-zero, which, in the absence of local order or pairwise correlations, is a key feature of SPT states. The signatures of a non-trivial string order in measurement data provide an interesting avenue for the use of machine learning to identify SPT phases~\cite{Sadoune:2024}.

\section{Quantum information properties of the Spin-1/2 Cluster State}

\begin{figure}[t]
    \centering
    \includegraphics[width=86mm]{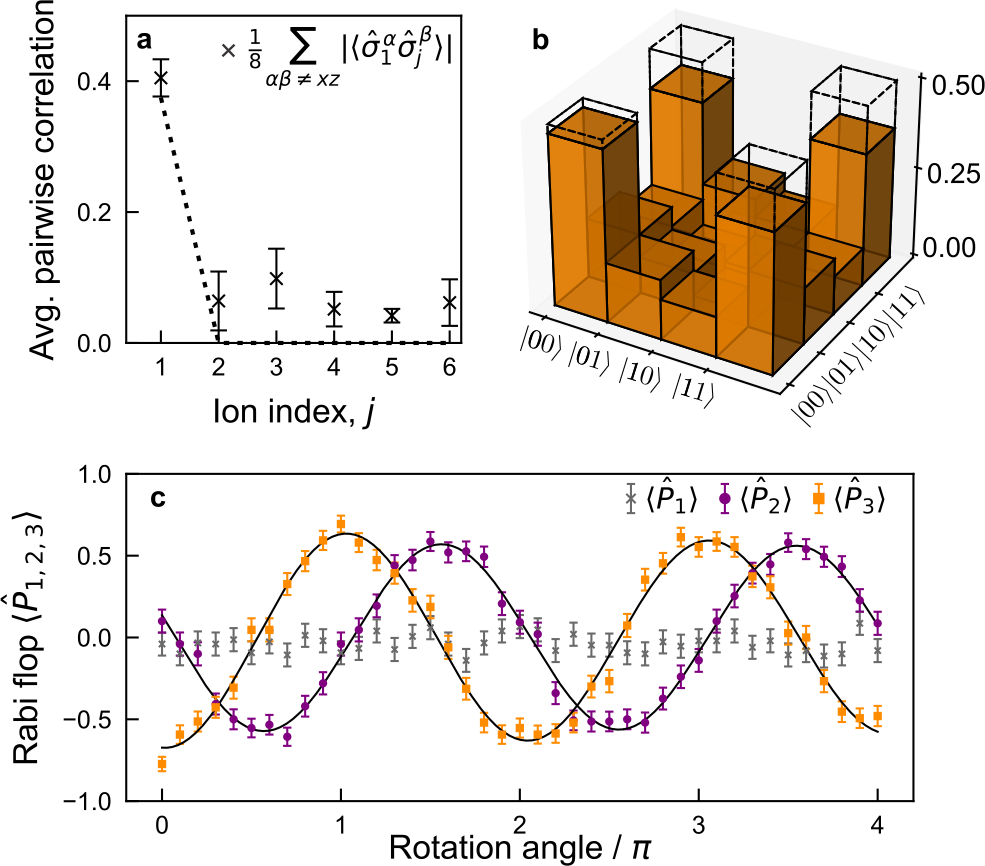}
    \caption{(a) The two-body Pauli correlations are measured for qubit-1 and qubit-$j$, $\langle\sxj{\alpha}{1}\sxj{\beta}{j}\rangle$ with $\alpha\beta\in\{x,y,z\}$. Excluding the measurement corresponding to a cluster state stabilizer, $\langle\sxj{x}{1}\sxj{z}{2}\rangle$, we average the deviation from zero for the remaining eight two-body correlations for each ion pair, $\tfrac{1}{8}\sum_{\alpha\beta\neq xz}|\langle\sxj{\alpha}{1}\sxj{\beta}{j}\rangle|$. The dashed line is a noise-free circuit simulation, which is identically zero for $j>1$. 
    (b) After post-selecting on the bulk qubit state $\ket{0000}$, the two-qubit state of the outer qubits is reconstructed using state tomography, giving an 81(3)\% fidelity compared to an ideal Bell state, $\tfrac{1}{\sqrt{2}}(\ket{00}+\ket{11})$ (transparent bars). (c) We drive Rabi flops on the effective qubit encoded within the cluster state using the unitary generated by $\hat{P}_1$ (see Eq.~\eqref{eq:cluster_P1}). The expectation values of the SU(2) operators are measured, $\hat{P}_1$ (gray), $\hat{P}_2$ (purple) and $\hat{P}_3$ (orange), and sinusoidal fits to $\langle\hat{P}_2\rangle$ and $\langle\hat{P}_3\rangle$ reveal a contrast of 0.58(3) and 0.67(3), respectively.}
    \label{fig:fig4}
\end{figure}

We now consider the qubit-based linear cluster state~\cite{Raussendorf:2001}, generated by applying controlled-Z entangling gates to neighboring sites in a linear chain of qubits in the initial state $\ket{+}$. Alternatively, the state can be understood as one of four degenerate ground states of the three-body interaction Hamiltonian with open boundary conditions,
\begin{equation}
\label{eq:ham_cluster}
    \hat{H}_{C} = - \sum_{i=2}^{N-1} \sxj{z}{i-1}\sxj{x}{i}\sxj{z}{i+1}.
\end{equation}
Notably, the linear cluster state is found to be an SPT state, protected by a $\mathbb{Z}_2\times\mathbb{Z}_2$ symmetry and closely related to the AKLT state, thus sharing many of its properties~\cite{Else:2012}. 
We experimentally generate a six-qubit linear cluster state using spin-1/2 trapped-ion qubits to study its quantum information and condensed matter properties, analogous to the AKLT state. This is achieved efficiently by entangling neighboring qubits using M{\o}lmer-S{\o}rensen gates, followed by a global $\RxT{y}{-\tfrac{\pi}{2}}$ rotation. Local rotations on the edge qubits (here $\RxT{x}{-\tfrac{\pi}{2}}$ on both qubits) prepare one of the four ground states. 

The ground states of $\hat{H}_C$ are those that attain $+1$ eigenvalues for all Hamiltonian terms and either $+1$ or $-1$ eigenvalues for the boundary terms $\sxj{x}{1}\sxj{z}{2}$ and \mbox{$\sxj{z}{N-1}\sxj{x}{N}$}~\cite{Raussendorf:2003}. Such states are known as stabilizer states and the collection of $N$ operators (Hamiltonian and boundary terms) that give eigenvalue $+1$ are known as the stabilizers of the state. Stabilizer states allow for particularly efficient verification~\cite{Ringbauer:2023} by sampling from the group of operators spanned by the stabilizers. For the six-qubit state, we obtain a fidelity of $\mathcal{F}=0.80(1)$.

Connecting to the analysis done for the AKLT state, we measure all single-body Pauli expectation values to have an average deviation from zero of 0.06(1), showing that the state is entangled and rotationally invariant, with no local order. Measuring the two-qubit expectation values reveals a finite correlation length, which, similar to the AKLT state, implies the existence of an energy gap. Indeed, excluding the two-body operator on the boundary that coincides with a stabilizer \mbox{$\langle \sxj{x}{1}\sxj{z}{2} \rangle$}, the average deviation from the expected zero for the remaining eight two-body expectation values is 0.06(3), shown in Fig.~\ref{fig:fig4}(a). Given the imperfect preparation fidelity of 0.80(1), this is well within the expectation of zero correlation length for the cluster state. Notably, despite displaying only very short-range correlations, the cluster state exhibits long-range order in analogy to the AKLT state. To reveal this order, we project the bulk spins onto one of the 16 possible outcomes. Each outcome leaves the edge qubits in one of four maximally entangled Bell states, with a fidelity of 81(3)\% (Fig.~\ref{fig:fig4}(b)). The above properties make the cluster state a blueprint for an ideal quantum repeater, where short-range correlations mean that it can be constructed locally, and the long-range order means that it can nonetheless connect distant parties. 

When restricted to the ground-state manifold, we now study the bulk-edge correspondence that is a characteristic of the SPT phase. Specifically, we identify an equivalence between two sets of operators acting globally on spins in the bulk, \mbox{$\{\mathbb{X}_\textrm{even},\,\mathbb{X}_\textrm{odd}\}$}, and two operators acting locally on the edges, \mbox{$\{\sxj{z}{1}\sxj{z}{N-1}\sxj{x}{N}, \, \sxj{x}{1}\sxj{z}{2}\sxj{z}{N}\}$} (see Appendix G for details). 
We show the equivalence of acting with either the bulk and edge operators on the ground state by comparing their effects on the stabilizer outcomes (Tab.~\ref{tab:bulkedge}). We observe that both bulk and edge operators preserve the ground-state manifold, as the average of the $2^4 = 16$ bulk operators generated by the stabilizers defined in $\hat{H}_C$ remains approximately constant. At the same time, the values of the two edge stabilizers oscillate with a contrast of about $90\%$ as the chain is driven between three of the four degenerate edge states. Based on these results, the bulk and edge operators equivalently permute the cluster state between states in the ground-state manifold. These operators could be used to manipulate a qubit located at the edge of the string, analogous to the AKLT chain.

\renewcommand{\arraystretch}{1.4}
\begin{table}
\begin{tabularx}{\linewidth} { 
   >{\centering\arraybackslash}X |
   >{\centering\arraybackslash}X 
   >{\centering\arraybackslash}c
   >{\centering\arraybackslash}c  }
    \hline
    Operator &
    $\sum_{i=2}^{5}\langle \sxj{z}{i-1}\sxj{x}{i}\sxj{z}{i+1} \rangle $  &
    $\langle \sxj{x}{1}\sxj{z}{2} \rangle$ &
    $\langle \sxj{z}{5}\sxj{x}{6} \rangle$ \\ \hline
     $\mathbb{I}_{2^N\times2^N}$& 0.80(2) & 0.91(2) & 0.89(3) \\ 
    $\mathbb{X}_\textrm{even}$& 0.82(2) & 0.93(2) & -0.89(3) \\
     $\sxj{z}{1}\sxj{z}{N-1}\sxj{x}{N}$ & 0.77(3) & 0.94(2) & -0.87(3) \\
     $\mathbb{X}_\textrm{odd}$& 0.80(2) & -0.92(2) & 0.88(3) \\ 
     $\sxj{x}{1}\sxj{z}{2}\sxj{z}{N}$ & 0.79(2) & -0.93(2) & 0.88(3) \\ \hline
\end{tabularx}
\caption{Bulk and edge stabilizer expectations after applying either the identity or one of the edge or bulk operators to the cluster state string, showing the preservation of the ground-state manifold and permutations through the degenerate edge states.}
\label{tab:bulkedge}
\end{table}

Finally, we can manipulate this effective qubit by identifying an SU(2) algebra generated by $\{\hat{P}_1, \hat{P}_2, \hat{P}_3\}$ (see Appendix G for full definitions). As with the AKLT state, we observe Rabi flops in Fig.~\ref{fig:fig4}(c) by applying the rotation generated by $\hat{P}_1$ and measuring the expectations of the three SU(2) observables.

\section{Discussion and Outlook}
The Haldane phase of the spin-1 Heisenberg chain is an archetypal model of SPT order and serves as a gateway for studying non-classical states of matter. Using a qudit-based quantum processor enables us to natively study such higher-spin systems efficiently, thereby extending the reach of quantum computers for studying condensed matter physics. A particularly interesting state within the Haldane phase is the well-known spin-1 AKLT state, which can be prepared efficiently through sequential coupling to an ancillary system, yet can be adiabatically connected to the full ground state. 

Beyond the AKLT state, the generation of states using sequential coupling via an ancilla qudit is a general methodology that can be applied to a range of matrix product states (MPS)~\cite{Cirac:2021}. 
Since the MPS links are mediated via the ancilla, the achievable bond dimension $D$ is given by the dimension $d$ of the ancilla qudit and can be conveniently controlled in the trapped-ion platform by choosing $d\geq D$. 
Using a single ancilla and qudits encoded in trapped $\Ca$ ions, we can already create other MPSs with bond dimensions up to $D=7$, with higher bond dimensions available using other ion species. 
Moreover, the order of the couplings between spins in the MPS is set by the application order of the sequential coupling unitaries, not the physical geometry of the quantum processor. Given the all-to-all coupling in our system, this means that arbitrary MPS geometries can be straightforwardly created, without the need to create complex computational mappings or physical trapping layouts. A key benefit of this approach is the efficient use of entanglement and the all-to-all connectivity of the trapped ion system, leading to a linear scaling with $N$.

The overarching goal, of course, is to extend the study of SPT phases beyond one spatial dimension to gain insights into realistic condensed matter systems and materials. Understanding and modeling the physics of 2D and 3D models, however, becomes increasingly challenging, with high expectations resting on quantum simulations to fill the gap.  Indeed, an analogous ``Haldane phase'' can be found for two-dimensional spin-1 grids, which similar to the 1D case can be prepared using a projected entangled pair state (PEPS). 
The sequential generation method used here is expected to enable the creation of this PEP state~\cite{Wei:2022}, as a starting point for studying 2D Haldane physics. In contrast to qubit-based computations, the native qudit implementation not only reduces the required number of quantum information carriers but also greatly simplifies the required interactions, and eliminates the need for complicated encoding and decoding steps. This hardware-efficient approach thus opens the door to a wide range of other applications in the quantum simulation of non-classical phases of matter. 

\section{Acknowledgements}
This research was funded by the European Union under the Horizon Europe Programme---Grant Agreements 101080086---NeQST, by the European Research Council (ERC, QUDITS, 101080086). Views and opinions expressed are however those of the author(s) only and do not necessarily reflect those of the European Union or the European Research Council Executive Agency. Neither the European Union nor the granting authority can be held responsible for them. We also acknowledge support by the Austrian Science Fund (FWF) through the SFB BeyondC (FWF Project No. F7109) and the EU-QUANTERA project TNiSQ (N-6001), and by the IQI GmbH. This project has received funding from the European Union’s Horizon 2020 research and innovation programme under the Marie Skłodowska‐Curie grant agreement No 801110 and the Austrian Federal Ministry of Education, Science, and Research (BMBWF). 
It reflects only the author's view, the EU Agency is not responsible for any use that may be made of the information it contains. G.K.B. acknowledges support from the Australian Research
Council Centre of Excellence for Engineered Quantum Systems (Grant No. CE 170100009).
E.R. acknowledges support from the BasQ strategy of the Department of Science, Universities, and Innovation of the Basque Government. E.R. is supported by the grant PID2021-126273NB-I00 funded by MCIN/AEI/ 10.13039/501100011033 and by ``ERDF A way of making Europe" and the Basque Government through Grant No. IT1470-22. This work was supported by the EU via QuantERA project T-NiSQ grant PCI2022-132984 funded by MCIN/AEI/10.13039/501100011033 and by the European Union ``NextGenerationEU''/PRTR. This work has been financially supported by the Ministry of Economic Affairs and Digital Transformation of the Spanish Government through the QUANTUM ENIA project called – Quantum Spain project, and by the European Union through the Recovery, Transformation, and Resilience Plan – NextGenerationEU within the framework of the Digital Spain 2026 Agenda. I. A. acknowledges support from the European Union’s Horizon Europe research and innovation programme under grant agreement No 101114305 (“MILLENION-SGA1” EU Project).
 
\section{Appendix}
\label{appendix}

\subsection{Experimental setup}

\begin{figure}[t]
    \centering
    \begin{center} 
    \includegraphics[scale=1]{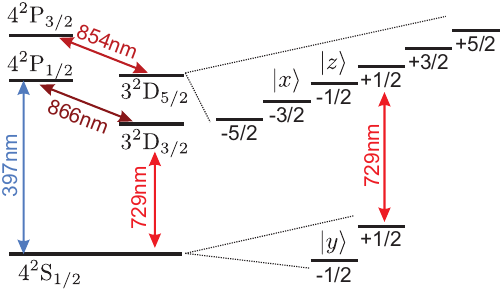}
    \end{center}
    \caption{The spin-1 qutrit is encoded in Zeeman sub-levels of $\level{S}{1/2}$ and $\level{D}{5/2}$ in $\Ca$, here labelled as \mbox{$\ket{x}, \ket{y}, \ket{z}$}. The qudit gates are performed on the quadrupole transition using a 729~nm laser. A diode laser at 397~nm is used for cooling, state preparation and detection; diode lasers at 866~nm and 854~nm are used for repumping.}
    \label{fig:energy_levels}
\end{figure}

Our experiment utilizes trapped $\Ca$ ions in a linear Paul trap. The qudits are encoded in the stable $\level{S}{1/2}$ and metastable $\level{D}{5/2}$, which are separated by an approximately 4.2~G magnetic field into two and six Zeeman sub-levels respectively (Fig.~\ref{fig:energy_levels}). Up to $d=7$ levels are available for computation, leaving one free level for population shelving during detection. The quadrupole transition is driven using a stabilized semiconductor laser at 729 nm. The dipole $\level{S}{1/2}
\leftrightarrow\level{P}{1/2}$ transition at 397~nm is used for optical pumping to $\level{S}{1/2}\ket{m_j=-1/2}$, Doppler cooling and state-selective fluorescence detection of states in the $\level{S}{1/2}$ manifold. Spin-1 local and entangling gates are decomposed into two-level gates using resonant and M{\o}lmer-S{\o}rensen (MS) operations. Detection is performed by successively detecting one of the $d$ states in the $\level{S}{1/2}$ manifold, with all other states shelved in the $\level{D}{5/2}$ manifold. Resonant $\pi$-pulses and polarization gradient cooling (PGC) are used between detections to transfer the shelved population to $\level{S}{1/2}$ and re-cool the ion string without affecting any unmeasured states~\cite{Ringbauer:2022}. In this work, we utilize strings of six to eight ions to create chains of up to five spin-1 particles plus one ancilla qubit.  

\subsection{The AKLT Model}

The AKLT Hamiltonian, defined in Eq.~\eqref{eq:ham_AKLT}, can be rewritten as a sum of projectors onto the spin-2 subspace on two adjacent lattice sites, \mbox{$\hat{H}_\textrm{AKLT}=\sum_{j} P_{S=2}\left( \Sj{j} + \Sj{j+1} \right)$}. To understand the AKLT state structure, we consider decomposing each spin-1 into a symmetric combination of two virtual spin-1/2 degrees of freedom, as shown on the left of Figure~\ref{fig:fig1} of the main text. Coupling the virtual particles on neighboring spin-1 sites into a spin-0 singlet ensures that we construct the chain exclusively using spin-1 sites and singlet coupling. Hence, each pair will have no overlap with the spin-2 subspace and contribute zero energy. Additionally, the Hamiltonian itself is a sum of projectors, guaranteeing a non-negative ground-state energy. Consequently, a state must reach a minimum of zero energy when the chain has no overlap with the spin-2 subspace. Therefore, this construction demonstrably leads to a ground state for the complete interacting model. 

The AKLT state can be written as a matrix product state (MPS)~\cite{Cirac:2021, Schuch:2011}. In the standard spin-1 basis, $\ket{s}\in\{\ket{+}, \ket{0}, \ket{-}\}$, the ground state can be expressed as the superposition of all possible spin combinations with prefactors calculated from the matrices $\hat{A}^{s_i}\in\{\hat{A}^{+}, \hat{A}^{0}, \hat{A}^{-}\}$, such that
\begin{equation}
    \ket{\Psi_\textrm{AKLT}} = \sum_{\{s\}} \textrm{Tr} [\hat{A}^{s_1} \dots \hat{A}^{s_N}] \ket{s_1\dots s_N}.
\end{equation}
The sum is over all possible spin states, $\ket{s}=\ket{s_1\dots s_N}$, in the length-$N$ chain, with the scalar prefactor of each term in the sum calculated from the MPS matrices. The dimension of these matrices sets the bond dimension of the state; for the AKLT state, the dimension is two, and the matrices are $2\times2$ matrices. They can be written as
\begin{equation}
    \hat{A}^{+} = \sqrt{\tfrac{2}{3}}\sx{+}, \, \hat{A}^{0} = -\sqrt{\tfrac{1}{3}}\sx{z}, \, \hat{A}^{-} = -\sqrt{\tfrac{2}{3}}\sx{-}, \,
\end{equation}
in terms of the Pauli spin-1/2 raising, $z$-component, and lowering operators respectively. 

To simplify the expression of the AKLT state, we utilize a symmetric spin-1 basis representation, which can be written in terms of the standard spin-1 basis as
\begin{equation}
    \ket{x} = \frac{\left(\ket{+} - \ket{-}\right)}{\sqrt{2}}, \, 
    \ket{y} = \frac{i\left(\ket{+} + \ket{-}\right)}{\sqrt{2}}, \, 
    \ket{z} = -\ket{0}.
\end{equation}
This basis is directly encoded into three energy levels of the $\Ca$ ions, as shown in Fig.~\ref{fig:energy_levels}. The corresponding spin-1 matrices are defined as
\begin{equation}
\begin{aligned}
    \tilde{\textbf{S}}
    &= \{ \tilde{S}^x, \tilde{S}^y, \tilde{S}^z \} \\
    &= \left\{
    \begin{pmatrix}
        0 & 0 & 0 \\
        0 & 0 & -i \\
        0 & i & 0
    \end{pmatrix}, 
    \begin{pmatrix}
        0 & 0 & i \\
        0 & 0 & 0 \\
        -i & 0 & 0
    \end{pmatrix}, \begin{pmatrix}
        0 & -i & 0 \\
        i & 0 & 0 \\
        0 & 0 & 0
    \end{pmatrix}
    \right\}
\end{aligned}
\end{equation}
such that $\tilde{S}^{a} |b\rangle = i \epsilon^{abc} |c\rangle$, with $\{a,b,c\} = \{x,y,z \}$ and $\epsilon^{abc}$ the Levi-Civita tensor. This basis is convenient for computation and is equivalent to the standard computational basis up to local rotations. With this basis, the MPS prefactors, $\hat{A}^{s_i}\in\{\hat{A}^{x}, \hat{A}^{y}, \hat{A}^{z}\}$, can be written simply in terms of the standard Pauli matrices as,
\begin{equation}
    \label{eq:AKLT_MPS_ops}
    \hat{A}^{x} = \tfrac{1}{\sqrt{3}}\sx{x}, \, \hat{A}^{y} = \tfrac{1}{\sqrt{3}}\sx{y},\, \hat{A}^{z} =\tfrac{1}{\sqrt{3}}\sx{z}.
\end{equation}

\subsection{Sequential generation algorithm for the AKLT state}

To generate the AKLT MPS as a sequentially generated state, each spin-1 particle is successively coupled to a shared spin-1/2 ancilla. The interaction between the ancilla and each spin-1 particle is derived from the set of MPS matrices, $\hat{A}^s$. If the spin-1 particles are initialized in the product state $\ket{s_0\dots s_0}$, then the interaction term between each spin-1 particle and the ancilla qubit can be calculated using the MPS matrices as
\begin{equation}
    \label{eq:gen_seq_unitary}
    \hat{U} = \sum_{s, \alpha, \beta} \hat{A}^s_{\beta\alpha} \ket{\beta, s}\bra{\alpha, s_0} 
\end{equation}
where the sum is over all possible input and output states of the ancilla system, \mbox{$\ket{\alpha}$} and \mbox{$\ket{\beta}$} respectively, and all possible output states of the spin-1 particle,  $\ket{s}$. Each term in the sum is multiplied by a scalar calculated from an element in the MPS operator matrix, $\hat{A}^s_{\beta\alpha}$, where the subscript identifies the row and column indices of the required matrix element. 

For the AKLT Hamiltonian in the symmetric spin-1 representation, we explicitly write out the terms with non-zero entries of $\hat{A}^s$ (Eq.~\eqref{eq:AKLT_MPS_ops}), when we initialize the spin-1 particles in $\ket{zz\dots z}$.
\begin{alignat}{3}
        \hat{U} =
        \,&\sx{x}_{10}&&\ket{\downarrow, x}\bra{\uparrow, z}
        + \sx{x}_{01} &&\ket{\uparrow, x}\bra{\downarrow, z} +\nonumber\\
        &\sx{y}_{10} &&\ket{\downarrow, y}\bra{\uparrow, z}
        + \sx{y}_{01}&&\ket{\uparrow, y}\bra{\downarrow, z} + \nonumber\\
        &\sx{z}_{00} &&\ket{\uparrow, z}\bra{\uparrow, z}
        + \sx{z}_{11}&&\ket{\downarrow, z}\bra{\downarrow, z}
\end{alignat}
From this, the two necessary mappings defining the unitary are given by,
\begin{equation}
    \begin{aligned}
        \hat{U}\ket{\uparrow, z}
        &= \ket{\downarrow, x} + i\ket{\downarrow, y} + \ket{\uparrow, z} \\
        \hat{U}\ket{\downarrow, z}
        &= \ket{\uparrow, x} - i\ket{\uparrow, y} -\ket{\downarrow, z}.
    \end{aligned}
\end{equation}
We can create a unitary that implements these two mappings with our physically available gate set using the BQSKit python package circuit synthesis tools~\cite{BQSKit}. The decomposition that we used in this work is shown in Table~\ref{tab:MPS_gate_decomp}, using two entangling gates and 25 local gates, which are implemented using 45 local two-level rotations in our experiment.

\begin{table}[t]
\begin{tabular}{ccccc}
    \textbf{Gate} & \textbf{Sublevels} &\textbf{ Rot. angle ($\theta$)} & \textbf{Axis ($\phi$)} & \textbf{Target}\\
    \hline
    $\Rx{z}$  &   (0, 1)   &   3.022   & -   &   A   \\ \hline
    $\Rx{z}$  &   (1, 2)   &   -3.222   & -   &   A   \\ \hline
    $\Rx{\phi}$  &   (1, 2)   &   0.245   & 1.414   &   A   \\ \hline
    $\Rx{\phi}$  &   (0, 1)   &   3.064   & 5.494   &   A   \\ \hline
    $\Rx{\phi}$  &   (1, 2)   &   0.743   & -0.539   &   A   \\ \hline
    $\textrm{MS}(\theta, \phi)$  &   (1, 2)   &   $\pi/2$   &     0   &   (A, S)   \\ \hline
    $\Rx{z}$  &   (0, 1)   &   1.444   & -   &   S   \\ \hline
    $\Rx{z}$  &   (1, 2)   &   -1.012   & -   &   S   \\ \hline
    $\Rx{\phi}$  &   (1, 2)   &   1.020   & 1.096   &   S   \\ \hline
    $\Rx{\phi}$  &   (0, 1)   &   2.218   & 3.244   &   S   \\ \hline
    $\Rx{\phi}$  &   (1, 2)   &   1.366   & 2.191   &   S   \\ \hline
    $\Rx{z}$  &   (0, 1)   &   5.748   & -   &   A   \\ \hline
    $\Rx{z}$  &   (1, 2)   &   6.254   & -   &   A   \\ \hline
    $\Rx{\phi}$  &   (1, 2)   &   1.521   & -1.293   &   A   \\ \hline
    $\Rx{\phi}$  &   (0, 1)   &   2.848   & 2.445   &   A  \\ \hline
    $\Rx{\phi}$  &   (1, 2)   &   0.219   & 4.322   &   A   \\ \hline
    $\textrm{MS}(\theta, \phi)$  &   (1, 2)   &   $\pi/2$   &     0   &   (A, S)   \\ \hline
    $\Rx{z}$  &   (0, 1)   &   3.047   & -   &   S   \\ \hline
    $\Rx{z}$  &   (1, 2)   &   1.511   & -   &   S   \\ \hline
    $\Rx{\phi}$  &   (1, 2)   &   1.488   & -2.107   &   S   \\ \hline
    $\Rx{\phi}$  &   (0, 1)   &   3.990   & 3.396   &   S   \\ \hline
    $\Rx{\phi}$  &   (1, 2)   &   0.961   & -2.615   &   S   \\ \hline
    $\Rx{z}$  &   (0, 1)   &   0.530   & -   &   A   \\ \hline
    $\Rx{z}$  &   (1, 2)   &   -2.678   & -   &   A   \\ \hline
    $\Rx{\phi}$  &   (1, 2)   &   0   & 0   &   A   \\ \hline
    $\Rx{\phi}$  &   (0, 1)   &   3.141   & 4.514   &   A   \\ \hline
    $\Rx{\phi}$  &   (1, 2)   &   0.708   & 1.571   &   A   \\ \hline
\end{tabular}
\caption{Decomposition of the sequential generation interaction in Eq.~\eqref{eq:AKLT_seq_unitary} using the BQSKit python package circuit synthesis tools~\cite{BQSKit}. Each rotation is performed on either the ancilla qubit (A) or the spin-1 particle (S), as shown in the final ``Target ion'' column. The angle and axis of rotation are both shown in radians.}
\label{tab:MPS_gate_decomp}  
\end{table}

\subsection{Energy measurement for the AKLT Hamiltonian}

To efficiently verify the engineered AKLT state, we measure the energy for the AKLT Hamiltonian,
\begin{equation}
\begin{aligned}
    \hat{H}_\textrm{AKLT} 
    &= \sum_{j=1}^{N-1} \left[
    \frac{1}{2} \SjT{j} \cdot \SjT{j+1} + \frac{1}{6}(\SjT{j} \cdot \SjT{j+1})^2 + \frac{1}{3} \right]\\
\end{aligned}
\end{equation}
known to be identically zero for the ground state. For each pair of neighbouring spins, the Hamiltonian terms can be divided into:
\begin{itemize}
\item  three linear terms of the form $\SjxT{j}{\alpha}\SjxT{j+1}{\alpha}$,
\item three quadratic terms of the form $(\SjxT{j}{\alpha}\SjxT{j+1}{\alpha})^2$, and 
\item three cross-terms, each of the form \mbox{$(\SjxT{j}{\alpha}\SjxT{j+1}{\alpha})(\SjxT{j}{\beta}\SjxT{j+1}{\beta}) + (\SjxT{j}{\beta}\SjxT{j+1}{\beta})(\SjxT{j}{\alpha}\SjxT{j+1}{\alpha})$},
\end{itemize}
with \mbox{$\alpha, \beta \in \{x, y, z\}$} and $\alpha \neq \beta$.

The energy $\langle \hat{H}_\textrm{AKLT} \rangle$ can be measured using nine measurements. The first three measurements record the expectation of the $x, y, $ or $z$ spin-1 operators on all particles simultaneously, $\langle \prod_{j=1}^N \SjxT{j}{\alpha} \rangle$. These measurements can be used to calculate the linear and quadratic contributions to the AKLT energy. The final three cross-terms are more complicated to measure but can each be measured in a single setting using an entangling gate. For each of the three cross-terms the measurement is performed twice; once measuring the neighboring pairs with $j$ odd, and once with $j$ even.

\subsection{Qutrit tomography with a mutually unbiased basis}

Performing state tomography in the standard Pauli basis returns an over-complete set of measurements. Instead, one can utilize a mutually unbiased basis (MUB) to reduce the number of measured bases, analogous to the scheme presented for qubits in Ref.~\cite{Stricker:2022}. As with standard tomography, the number of measurements will scale exponentially, however, by reducing the base of the exponent we can extend to measuring larger than usual qutrit states, here measuring states with up to, but not limited to, $N=4$ qutrits. A total of nine basis states per qutrit are required for informational completeness, which can be constructed using four measurements, each with three possible outcomes. As a result, the number of measurements scales as $4^N$. The basis set is defined as
\setstacktabbedgap{20pt}
\begin{equation}
\begin{aligned}
    &\begin{pmatrix} 
    \begin{array}{w{c}{\widthof{$\omega^2$}}
                  w{c}{\widthof{$\omega^2$}}
                  w{c}{\widthof{$\omega^2$}}}
        1 & 0 & 0 \\
        0 & 1 & 0 \\
        0 & 0 & 1
    \end{array}
    \end{pmatrix}, 
    &&\frac{1}{\sqrt{3}}
    \begin{pmatrix}
    \begin{array}{w{c}{\widthof{$\omega^2$}}
                  w{c}{\widthof{$\omega^2$}}
                  w{c}{\widthof{$\omega^2$}}}
        1 & 1 & 1 \\
        1 & \omega & \omega^2 \\
        1 & \omega^2 & \omega
    \end{array}
    \end{pmatrix}, \\
    \frac{1}{\sqrt{3}}
    &\begin{pmatrix}
    \begin{array}{w{c}{\widthof{$\omega^2$}}
                  w{c}{\widthof{$\omega^2$}}
                  w{c}{\widthof{$\omega^2$}}}
        1 & 1 & 1 \\
        \omega^2 & 1 & \omega \\
        \omega^2 & \omega & 1
    \end{array}
    \end{pmatrix}, 
    &&\frac{1}{\sqrt{3}}
    \begin{pmatrix}
    \begin{array}{w{c}{\widthof{$\omega^2$}}
                  w{c}{\widthof{$\omega^2$}}
                  w{c}{\widthof{$\omega^2$}}}
        1 & 1 & 1 \\
        \omega & \omega^2 & 1 \\
        \omega & 1 & \omega^2
    \end{array}
    \end{pmatrix}
\end{aligned}
\end{equation}
where $\omega = e^{\frac{2i\pi}{3}}$. By measuring the state in the $4^N$ settings generated by these four bases, we can reconstruct the $N$-qutrit state using linear inversion or maximum likelihood estimation.

\subsection{Rabi flops of the fractionalized edge}

\begin{figure}[h]
    \centering
    \begin{center} 
    \includegraphics[scale=1]{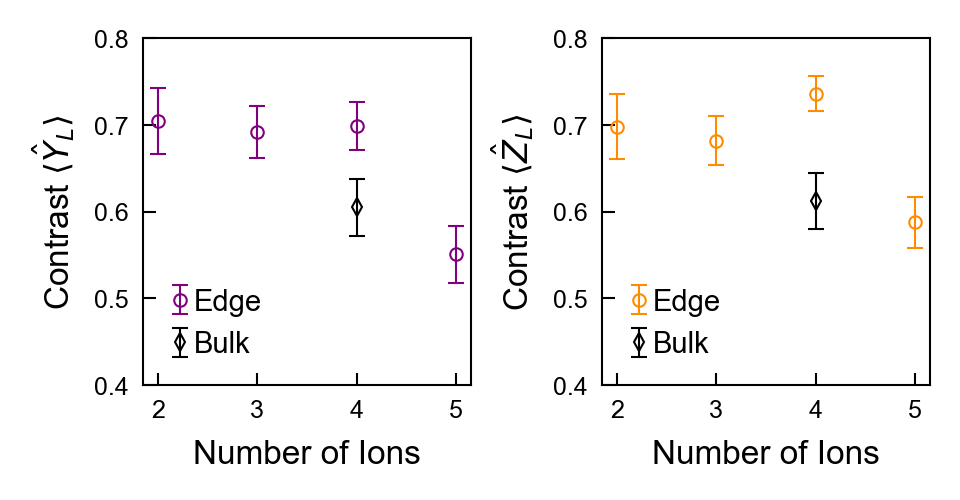}
    \end{center}
    \caption{Measuring the SU(2) Rabi flop contrasts for AKLT chains of different lengths, $N$. We fit a decaying sine curve to the AKLT ground-state manifold Rabi flops, as in Fig.~\ref{fig:fig2} of the main text. This is done for both the edge-driven rotations generated by $\hat{X}_L$ (purple and orange circles), and the bulk-driven global rotation (black diamonds). For the latter, the data is only taken for length $N=4$ as a comparison to the edge-driven data. The contrasts are plotted for the expectation values $\langle\hat{Y}_L\rangle$ (left) and $\langle\hat{Z}_L\rangle$ (right). Error bars are one standard deviation calculated from Monte Carlo resampling.}
    \label{fig:rabi_contrasts}
\end{figure}

In Fig.~\ref{fig:rabi_contrasts} we compare the Rabi flop contrasts when driving between the AKLT ground-state manifold states for increasing chain length, $N$. The SU(2) algebra can be demonstrated equivalently using a global rotation on the bulk, \mbox{$ e^{-i\theta\sum_{j=1}^N \SjxT{j}{x}/2}$}, or using an edge-driven rotation generated by the operator $\hat{X}_L$ defined in Eq.~\eqref{eq:logic_x}. We fit the contrasts of the Rabi flops when measuring the expectations of the operators $\hat{Y}_L$ and $\hat{Z}_L$, defined in Eqs.~\eqref{eq:logic_y}-\eqref{eq:logic_z}. The bulk-driven Rabi flops were measured for only one chain length, $N=4$, to show equivalence in the behavior between the bulk- and edge operators on the AKLT ground-state manifold.

For chains of length $N=2-4$, the contrast of the edge-driven Rabi flops remains approximately constant around 69\%-74\%. The drop in Rabi contrast for $N=5$ sites is explained by a change in the experimental setup. All AKLT states with $N\leq4$ were produced with six $\Ca$ ions in the trap. The $N=5$ chain required eight $\Ca$ ions in the trap to minimize cross-talk between the qutrits. The increase in the number of ions is known to reduce the fidelity of the individual gates~\cite{Ringbauer:2022}, hence reducing the quality of the fractionalized edge qubit.

\subsection{The Spin-1/2 Cluster State}

The cluster state, $\ket{\Psi_{C}}$, is defined as one of the ground states of the Hamiltonian built from the following three-body operators,
\begin{equation}
    \hat{H}_{C} = - \sum_{i=2}^{N-1} \sxj{z}{i-1}\sxj{x}{i}\sxj{z}{i+1}.
\end{equation}
The cluster state is a \textit{stabilizer state}~\cite{Raussendorf:2003}, stabilized by each of the component terms in $H_C$, such that \mbox{$\sxj{z}{i-1} \sxj{x}{i}\sxj{z}{i+1}\ket{\Psi_{C}} = +\ket{\Psi_{C}}$}. Equivalently, we can identify each stabilizer term as equivalent to the identity operator when restricted to the ground-state manifold. We are then left with two degenerate degrees of freedom due to the open boundary conditions, resulting in a four-fold degenerate ground-state subspace.

If the total number of sites $N$ is even, then $H_{C}$ is symmetric under the $\mathbb{Z}_{2} \times \mathbb{Z}_{2} $ group generated by the non-local string operators
\begin{equation}
\begin{split}
    \mathbb{X}_{odd} &=
    \sxj{x}{1}\Idj{2}\sxj{x}{3}\Idj{4}\sxj{x}{5}
    \cdots \sxj{x}{N-1}\Idj{N} \\
    \mathbb{X}_{even} &= 
    \Idj{1}\sxj{x}{2}\Idj{3}\sxj{x}{4}\Idj{5}\sxj{x}{6}
    \cdots \Idj{N-1}\sxj{x}{N}.
\end{split}
\end{equation}
These are the ``bulk'' operators described in the main text, as they act across the entire string. To identify an equivalence between the bulk operators and an operator localized at the edge, we use the bulk stabilizer operators. Multiplying every second term in the Hamiltonian \mbox{($i=2,4,\dots, N-2)$} leads to the following condition in the ground-state manifold, 
\begin{align}
    (\sxj{z}{1}\sxj{x}{2}\sxj{z}{3})
    (\sxj{z}{3}\sxj{x}{4}\sxj{z}{5}) \cdots (\sxj{z}{N-2}\sxj{x}{N-1}\sxj{z}{N}) &= \mathbb{I}_{2^N\times 2^N} \nonumber \\
    \sxj{z}{1}\sxj{x}{2}\Idj{3}\sxj{x}{4}\Idj{5}\sxj{x}{6} \cdots \sxj{x}{N-2} \sxj{z}{N-1}\Idj{N} &= \mathbb{I}_{2^N\times 2^N},
\end{align}
where $\mathbb{I}_{2^N\times 2^N}$ is the identity operator acting on the entire string. We can rewrite this using the bulk operators as \mbox{$\sxj{z}{1}\mathbb{X}_{even} \sxj{z}{N-1}\sxj{x}{N} = \mathbb{I}_{2^N\times 2^N}$},
or equivalently 
\begin{equation}
    \mathbb{X}_{even}|_\textrm{GS} = \sxj{z}{1}\sxj{z}{N-1}\sxj{x}{N},
\end{equation}
where GS restricts the operator to the ground-state manifold. Therefore, despite $\mathbb{X}_{even}$ being a bulk operator acting on the whole string, it can be redefined as an action localized on the edges when restricted to the ground-state subspace. Similarly, $\mathbb{X}_{odd}$ can be redefined on the ground-state manifold as 
\begin{equation}
    \mathbb{X}_{odd}|_\textrm{GS} = \sxj{x}{1}\sxj{z}{2}\sxj{z}{N}
\end{equation}
by multiplying every alternate bulk stabilizer term from $i=3$. 

We can decompose the operator into two components acting on the left and right edges as \mbox{$\mathbb{X}_{even}^{L} = \sxj{z}{1}$} and \mbox{$\mathbb{X}_{even}^{R} = \sxj{z}{N-1}\sxj{x}{N}$} respectively. Similarly, $\mathbb{X}_{odd}$ can be decomposed in the ground-state manifold as \mbox{$\mathbb{X}_{odd}^{L}= \sxj{x}{1}\sxj{z}{2}$} and \mbox{$\mathbb{X}_{odd}^{R} = \sxj{z}{N}$}. It can be straightforwardly shown that $\mathbb{X}_{odd}^{L}$ and $\mathbb{X}_{even}^{L}$ are anti-commuting symmetries, proving that the cluster model has degenerate edge modes. 

From \mbox{$\mathbb{X}_{odd}\mathbb{X}_{even} = \mathbb{X}_{even}\mathbb{X}_{odd}$} one can derive that \mbox{$\mathbb{X}_{odd}^{L} \mathbb{X}_{even}^{L} = e^{i\alpha} \mathbb{X}_{even}^{L} \mathbb{X}_{odd}^{L}$}. Furthermore, using \mbox{$\mathbb{X}_{odd}^{2}= 1$} one can show that \mbox{$e^{i\alpha} = \pm 1$}, indeed labelling the projective representations of $\mathbb{Z}_{2} \times \mathbb{Z}_{2}$. Thus as long as the correlation length is finite, the edges have a well-defined degeneracy. Hence the cluster model is an SPT phase protected by $\mathbb{Z}_{2} \times \mathbb{Z}_{2}$. The $\mathbb{Z}_{2} \times \mathbb{Z}_{2}$ symmetry group can be written in terms of an SU(2) algebra generated by the symmetry operators,
\begin{align}
    \label{eq:cluster_P1}
    \hat{P}_1 &= \Idj{1}\sxj{x}{2}\Idj{3}\sxj{x}{4}\dots\sxj{x}{N-2}\sxj{z}{N-1}\Idj{N} \\
    \label{eq:cluster_P2}
    \hat{P}_2 &= -\sxj{x}{1}\sxj{x}{2}\sxj{x}{3}\sxj{x}{4}\dots\sxj{x}{N-2}\sxj{y}{N-1}\sxj{z}{N} \\
    \label{eq:cluster_P3}
    \hat{P}_3 &= \sxj{x}{1}\Idj{2}\sxj{x}{3}\Idj{4}\dots\Idj{N-2}\sxj{x}{N-1}\sxj{z}{N}.
\end{align}
It can be shown that \mbox{$[\hat{P}_i, \hat{P}_j] = 2i\epsilon_{ijk}\hat{P}_k$}, with the $\epsilon_{ijk}$ the Levi-Civita symbol. We can manipulate the SU(2) algebra within the cluster state by driving Rabi flops generated by $\hat{P}_1$, as shown in Fig.~\ref{fig:fig4} of the main text.

\bibliography{main}

\begin{thebibliography}{48}%
\makeatletter
\providecommand \@ifxundefined [1]{%
 \@ifx{#1\undefined}
}%
\providecommand \@ifnum [1]{%
 \ifnum #1\expandafter \@firstoftwo
 \else \expandafter \@secondoftwo
 \fi
}%
\providecommand \@ifx [1]{%
 \ifx #1\expandafter \@firstoftwo
 \else \expandafter \@secondoftwo
 \fi
}%
\providecommand \natexlab [1]{#1}%
\providecommand \enquote  [1]{``#1''}%
\providecommand \bibnamefont  [1]{#1}%
\providecommand \bibfnamefont [1]{#1}%
\providecommand \citenamefont [1]{#1}%
\providecommand \href@noop [0]{\@secondoftwo}%
\providecommand \href [0]{\begingroup \@sanitize@url \@href}%
\providecommand \@href[1]{\@@startlink{#1}\@@href}%
\providecommand \@@href[1]{\endgroup#1\@@endlink}%
\providecommand \@sanitize@url [0]{\catcode `\\12\catcode `\$12\catcode
  `\&12\catcode `\#12\catcode `\^12\catcode `\_12\catcode `\%12\relax}%
\providecommand \@@startlink[1]{}%
\providecommand \@@endlink[0]{}%
\providecommand \url  [0]{\begingroup\@sanitize@url \@url }%
\providecommand \@url [1]{\endgroup\@href {#1}{\urlprefix }}%
\providecommand \urlprefix  [0]{URL }%
\providecommand \Eprint [0]{\href }%
\providecommand \doibase [0]{http://dx.doi.org/}%
\providecommand \selectlanguage [0]{\@gobble}%
\providecommand \bibinfo  [0]{\@secondoftwo}%
\providecommand \bibfield  [0]{\@secondoftwo}%
\providecommand \translation [1]{[#1]}%
\providecommand \BibitemOpen [0]{}%
\providecommand \bibitemStop [0]{}%
\providecommand \bibitemNoStop [0]{.\EOS\space}%
\providecommand \EOS [0]{\spacefactor3000\relax}%
\providecommand \BibitemShut  [1]{\csname bibitem#1\endcsname}%
\let\auto@bib@innerbib\@empty
\bibitem [{\citenamefont {Kumar}\ \emph {et~al.}(2021)\citenamefont {Kumar},
  \citenamefont {Guin}, \citenamefont {Manna}, \citenamefont {Shekhar},\ and\
  \citenamefont {Felser}}]{Kumar:2021}%
  \BibitemOpen
  \bibfield  {author} {\bibinfo {author} {\bibfnamefont {Nitesh}\ \bibnamefont
  {Kumar}}, \bibinfo {author} {\bibfnamefont {Satya~N.}\ \bibnamefont {Guin}},
  \bibinfo {author} {\bibfnamefont {Kaustuv}\ \bibnamefont {Manna}}, \bibinfo
  {author} {\bibfnamefont {Chandra}\ \bibnamefont {Shekhar}}, \ and\ \bibinfo
  {author} {\bibfnamefont {Claudia}\ \bibnamefont {Felser}},\ }\bibfield
  {title} {\enquote {\bibinfo {title} {Topological quantum materials from the
  viewpoint of chemistry},}\ }\href {\doibase 10.1021/acs.chemrev.0c00732}
  {\bibfield  {journal} {\bibinfo  {journal} {Chemical Reviews}\ }\textbf
  {\bibinfo {volume} {121}},\ \bibinfo {pages} {2780--2815} (\bibinfo {year}
  {2021})}\BibitemShut {NoStop}%
\bibitem [{\citenamefont {Huang}\ \emph {et~al.}(2017)\citenamefont {Huang},
  \citenamefont {Xu}, \citenamefont {Wang},\ and\ \citenamefont
  {Duan}}]{Huang:2017}%
  \BibitemOpen
  \bibfield  {author} {\bibinfo {author} {\bibfnamefont {Huaqing}\ \bibnamefont
  {Huang}}, \bibinfo {author} {\bibfnamefont {Yong}\ \bibnamefont {Xu}},
  \bibinfo {author} {\bibfnamefont {Jianfeng}\ \bibnamefont {Wang}}, \ and\
  \bibinfo {author} {\bibfnamefont {Wenhui}\ \bibnamefont {Duan}},\ }\bibfield
  {title} {\enquote {\bibinfo {title} {Emerging topological states in
  quasi-two-dimensional materials},}\ }\href {\doibase
  https://doi.org/10.1002/wcms.1296} {\bibfield  {journal} {\bibinfo  {journal}
  {WIREs Computational Molecular Science}\ }\textbf {\bibinfo {volume} {7}},\
  \bibinfo {pages} {e1296} (\bibinfo {year} {2017})}\BibitemShut {NoStop}%
\bibitem [{\citenamefont {Jin}\ \emph {et~al.}(2023)\citenamefont {Jin},
  \citenamefont {Jiang}, \citenamefont {Sethi},\ and\ \citenamefont
  {Liu}}]{Jin:2023}%
  \BibitemOpen
  \bibfield  {author} {\bibinfo {author} {\bibfnamefont {Kyung-Hwan}\
  \bibnamefont {Jin}}, \bibinfo {author} {\bibfnamefont {Wei}\ \bibnamefont
  {Jiang}}, \bibinfo {author} {\bibfnamefont {Gurjyot}\ \bibnamefont {Sethi}},
  \ and\ \bibinfo {author} {\bibfnamefont {Feng}\ \bibnamefont {Liu}},\
  }\bibfield  {title} {\enquote {\bibinfo {title} {Topological quantum devices:
  a review},}\ }\href {\doibase 10.1039/D3NR01288C} {\bibfield  {journal}
  {\bibinfo  {journal} {Nanoscale}\ }\textbf {\bibinfo {volume} {15}},\
  \bibinfo {pages} {12787--12817} (\bibinfo {year} {2023})}\BibitemShut
  {NoStop}%
\bibitem [{\citenamefont {Miller}\ and\ \citenamefont
  {Miyake}(2015)}]{Miller:2015}%
  \BibitemOpen
  \bibfield  {author} {\bibinfo {author} {\bibfnamefont {Jacob}\ \bibnamefont
  {Miller}}\ and\ \bibinfo {author} {\bibfnamefont {Akimasa}\ \bibnamefont
  {Miyake}},\ }\bibfield  {title} {\enquote {\bibinfo {title} {Resource quality
  of a symmetry-protected topologically ordered phase for quantum
  computation},}\ }\href {\doibase 10.1103/PhysRevLett.114.120506} {\bibfield
  {journal} {\bibinfo  {journal} {Phys. Rev. Lett.}\ }\textbf {\bibinfo
  {volume} {114}},\ \bibinfo {pages} {120506} (\bibinfo {year}
  {2015})}\BibitemShut {NoStop}%
\bibitem [{\citenamefont {Bartlett}\ \emph {et~al.}(2017)\citenamefont
  {Bartlett}, \citenamefont {Brennen},\ and\ \citenamefont
  {Miyake}}]{Bartlett:2018}%
  \BibitemOpen
  \bibfield  {author} {\bibinfo {author} {\bibfnamefont {Stephen~D}\
  \bibnamefont {Bartlett}}, \bibinfo {author} {\bibfnamefont {Gavin~K}\
  \bibnamefont {Brennen}}, \ and\ \bibinfo {author} {\bibfnamefont {Akimasa}\
  \bibnamefont {Miyake}},\ }\bibfield  {title} {\enquote {\bibinfo {title}
  {Robust symmetry-protected metrology with the haldane phase},}\ }\href
  {\doibase 10.1088/2058-9565/aa9c56} {\bibfield  {journal} {\bibinfo
  {journal} {Quantum Sci. Tech.}\ }\textbf {\bibinfo {volume} {3}},\ \bibinfo
  {pages} {014010} (\bibinfo {year} {2017})}\BibitemShut {NoStop}%
\bibitem [{\citenamefont {Pollmann}\ \emph {et~al.}(2012)\citenamefont
  {Pollmann}, \citenamefont {Berg}, \citenamefont {Turner},\ and\ \citenamefont
  {Oshikawa}}]{Pollman:2012_feb}%
  \BibitemOpen
  \bibfield  {author} {\bibinfo {author} {\bibfnamefont {Frank}\ \bibnamefont
  {Pollmann}}, \bibinfo {author} {\bibfnamefont {Erez}\ \bibnamefont {Berg}},
  \bibinfo {author} {\bibfnamefont {Ari~M.}\ \bibnamefont {Turner}}, \ and\
  \bibinfo {author} {\bibfnamefont {Masaki}\ \bibnamefont {Oshikawa}},\
  }\bibfield  {title} {\enquote {\bibinfo {title} {Symmetry protection of
  topological phases in one-dimensional quantum spin systems},}\ }\href
  {\doibase 10.1103/PhysRevB.85.075125} {\bibfield  {journal} {\bibinfo
  {journal} {Phys. Rev. B}\ }\textbf {\bibinfo {volume} {85}},\ \bibinfo
  {pages} {075125} (\bibinfo {year} {2012})}\BibitemShut {NoStop}%
\bibitem [{\citenamefont {Pollmann}\ and\ \citenamefont
  {Turner}(2012)}]{Pollman:2012_sep}%
  \BibitemOpen
  \bibfield  {author} {\bibinfo {author} {\bibfnamefont {Frank}\ \bibnamefont
  {Pollmann}}\ and\ \bibinfo {author} {\bibfnamefont {Ari~M.}\ \bibnamefont
  {Turner}},\ }\bibfield  {title} {\enquote {\bibinfo {title} {Detection of
  symmetry-protected topological phases in one dimension},}\ }\href {\doibase
  10.1103/PhysRevB.86.125441} {\bibfield  {journal} {\bibinfo  {journal} {Phys.
  Rev. B}\ }\textbf {\bibinfo {volume} {86}},\ \bibinfo {pages} {125441}
  (\bibinfo {year} {2012})}\BibitemShut {NoStop}%
\bibitem [{\citenamefont {Verresen}\ \emph {et~al.}(2017)\citenamefont
  {Verresen}, \citenamefont {Moessner},\ and\ \citenamefont
  {Pollmann}}]{Pollman:2017}%
  \BibitemOpen
  \bibfield  {author} {\bibinfo {author} {\bibfnamefont {Ruben}\ \bibnamefont
  {Verresen}}, \bibinfo {author} {\bibfnamefont {Roderich}\ \bibnamefont
  {Moessner}}, \ and\ \bibinfo {author} {\bibfnamefont {Frank}\ \bibnamefont
  {Pollmann}},\ }\bibfield  {title} {\enquote {\bibinfo {title}
  {One-dimensional symmetry protected topological phases and their
  transitions},}\ }\href {\doibase 10.1103/PhysRevB.96.165124} {\bibfield
  {journal} {\bibinfo  {journal} {Phys. Rev. B}\ }\textbf {\bibinfo {volume}
  {96}},\ \bibinfo {pages} {165124} (\bibinfo {year} {2017})}\BibitemShut
  {NoStop}%
\bibitem [{\citenamefont {Ellison}\ \emph {et~al.}(2021)\citenamefont
  {Ellison}, \citenamefont {Kato}, \citenamefont {Liu},\ and\ \citenamefont
  {Hsieh}}]{Ellison:2021}%
  \BibitemOpen
  \bibfield  {author} {\bibinfo {author} {\bibfnamefont {Tyler~D.}\
  \bibnamefont {Ellison}}, \bibinfo {author} {\bibfnamefont {Kohtaro}\
  \bibnamefont {Kato}}, \bibinfo {author} {\bibfnamefont {Zi-Wen}\ \bibnamefont
  {Liu}}, \ and\ \bibinfo {author} {\bibfnamefont {Timothy~H.}\ \bibnamefont
  {Hsieh}},\ }\bibfield  {title} {\enquote {\bibinfo {title}
  {Symmetry-protected sign problem and magic in quantum phases of matter},}\
  }\href {\doibase 10.22331/q-2021-12-28-612} {\bibfield  {journal} {\bibinfo
  {journal} {Quantum}\ }\textbf {\bibinfo {volume} {5}},\ \bibinfo {pages}
  {612} (\bibinfo {year} {2021})}\BibitemShut {NoStop}%
\bibitem [{\citenamefont {Lanyon}\ \emph {et~al.}(2011)\citenamefont {Lanyon},
  \citenamefont {Hempel}, \citenamefont {Nigg}, \citenamefont {Müller},
  \citenamefont {Gerritsma}, \citenamefont {Zähringer}, \citenamefont
  {Schindler}, \citenamefont {Barreiro}, \citenamefont {Rambach}, \citenamefont
  {Kirchmair}, \citenamefont {Hennrich}, \citenamefont {Zoller}, \citenamefont
  {Blatt},\ and\ \citenamefont {Roos}}]{Lanyon:2011}%
  \BibitemOpen
  \bibfield  {author} {\bibinfo {author} {\bibfnamefont {B.~P.}\ \bibnamefont
  {Lanyon}}, \bibinfo {author} {\bibfnamefont {C.}~\bibnamefont {Hempel}},
  \bibinfo {author} {\bibfnamefont {D.}~\bibnamefont {Nigg}}, \bibinfo {author}
  {\bibfnamefont {M.}~\bibnamefont {Müller}}, \bibinfo {author} {\bibfnamefont
  {R.}~\bibnamefont {Gerritsma}}, \bibinfo {author} {\bibfnamefont
  {F.}~\bibnamefont {Zähringer}}, \bibinfo {author} {\bibfnamefont
  {P.}~\bibnamefont {Schindler}}, \bibinfo {author} {\bibfnamefont {J.~T.}\
  \bibnamefont {Barreiro}}, \bibinfo {author} {\bibfnamefont {M.}~\bibnamefont
  {Rambach}}, \bibinfo {author} {\bibfnamefont {G.}~\bibnamefont {Kirchmair}},
  \bibinfo {author} {\bibfnamefont {M.}~\bibnamefont {Hennrich}}, \bibinfo
  {author} {\bibfnamefont {P.}~\bibnamefont {Zoller}}, \bibinfo {author}
  {\bibfnamefont {R.}~\bibnamefont {Blatt}}, \ and\ \bibinfo {author}
  {\bibfnamefont {C.~F.}\ \bibnamefont {Roos}},\ }\bibfield  {title} {\enquote
  {\bibinfo {title} {Universal digital quantum simulation with trapped ions},}\
  }\href {\doibase 10.1126/science.1208001} {\bibfield  {journal} {\bibinfo
  {journal} {Science}\ }\textbf {\bibinfo {volume} {334}},\ \bibinfo {pages}
  {57--61} (\bibinfo {year} {2011})}\BibitemShut {NoStop}%
\bibitem [{\citenamefont {Meth}\ \emph {et~al.}(2022)\citenamefont {Meth},
  \citenamefont {Kuzmin}, \citenamefont {van Bijnen}, \citenamefont {Postler},
  \citenamefont {Stricker}, \citenamefont {Blatt}, \citenamefont {Ringbauer},
  \citenamefont {Monz}, \citenamefont {Silvi},\ and\ \citenamefont
  {Schindler}}]{Meth:2022}%
  \BibitemOpen
  \bibfield  {author} {\bibinfo {author} {\bibfnamefont {Michael}\ \bibnamefont
  {Meth}}, \bibinfo {author} {\bibfnamefont {Viacheslav}\ \bibnamefont
  {Kuzmin}}, \bibinfo {author} {\bibfnamefont {Rick}\ \bibnamefont {van
  Bijnen}}, \bibinfo {author} {\bibfnamefont {Lukas}\ \bibnamefont {Postler}},
  \bibinfo {author} {\bibfnamefont {Roman}\ \bibnamefont {Stricker}}, \bibinfo
  {author} {\bibfnamefont {Rainer}\ \bibnamefont {Blatt}}, \bibinfo {author}
  {\bibfnamefont {Martin}\ \bibnamefont {Ringbauer}}, \bibinfo {author}
  {\bibfnamefont {Thomas}\ \bibnamefont {Monz}}, \bibinfo {author}
  {\bibfnamefont {Pietro}\ \bibnamefont {Silvi}}, \ and\ \bibinfo {author}
  {\bibfnamefont {Philipp}\ \bibnamefont {Schindler}},\ }\bibfield  {title}
  {\enquote {\bibinfo {title} {Probing phases of quantum matter with an
  ion-trap tensor-network quantum eigensolver},}\ }\href {\doibase
  10.1103/PhysRevX.12.041035} {\bibfield  {journal} {\bibinfo  {journal} {Phys.
  Rev. X}\ }\textbf {\bibinfo {volume} {12}},\ \bibinfo {pages} {041035}
  (\bibinfo {year} {2022})}\BibitemShut {NoStop}%
\bibitem [{\citenamefont {Haldane}(1983{\natexlab{a}})}]{Haldane:1983}%
  \BibitemOpen
  \bibfield  {author} {\bibinfo {author} {\bibfnamefont {F.D.M.}\ \bibnamefont
  {Haldane}},\ }\bibfield  {title} {\enquote {\bibinfo {title} {Continuum
  dynamics of the 1-d heisenberg antiferromagnet: Identification with the o(3)
  nonlinear sigma model},}\ }\href {\doibase
  https://doi.org/10.1016/0375-9601(83)90631-X} {\bibfield  {journal} {\bibinfo
   {journal} {Phys. Lett. A}\ }\textbf {\bibinfo {volume} {93}},\ \bibinfo
  {pages} {464--468} (\bibinfo {year} {1983}{\natexlab{a}})}\BibitemShut
  {NoStop}%
\bibitem [{\citenamefont {Haldane}(1983{\natexlab{b}})}]{Haldane:1983_apr}%
  \BibitemOpen
  \bibfield  {author} {\bibinfo {author} {\bibfnamefont {F.~D.~M.}\
  \bibnamefont {Haldane}},\ }\bibfield  {title} {\enquote {\bibinfo {title}
  {Nonlinear field theory of large-spin heisenberg antiferromagnets:
  Semiclassically quantized solitons of the one-dimensional easy-axis n\'eel
  state},}\ }\href {\doibase 10.1103/PhysRevLett.50.1153} {\bibfield  {journal}
  {\bibinfo  {journal} {Phys. Rev. Lett.}\ }\textbf {\bibinfo {volume} {50}},\
  \bibinfo {pages} {1153--1156} (\bibinfo {year}
  {1983}{\natexlab{b}})}\BibitemShut {NoStop}%
\bibitem [{\citenamefont {Haldane}(2016)}]{Haldane:2016}%
  \BibitemOpen
  \bibfield  {author} {\bibinfo {author} {\bibfnamefont {F.~D.~M.}\
  \bibnamefont {Haldane}},\ }\href@noop {} {\enquote {\bibinfo {title} {Ground
  state properties of antiferromagnetic chains with unrestricted spin: Integer
  spin chains as realisations of the o(3) non-linear sigma model},}\ }
  (\bibinfo {year} {2016}),\ \Eprint {http://arxiv.org/abs/1612.00076}
  {arXiv:1612.00076} \BibitemShut {NoStop}%
\bibitem [{\citenamefont {Affleck}\ \emph {et~al.}(1987)\citenamefont
  {Affleck}, \citenamefont {Kennedy}, \citenamefont {Lieb},\ and\ \citenamefont
  {Tasaki}}]{Affleck:1987}%
  \BibitemOpen
  \bibfield  {author} {\bibinfo {author} {\bibfnamefont {Ian}\ \bibnamefont
  {Affleck}}, \bibinfo {author} {\bibfnamefont {Tom}\ \bibnamefont {Kennedy}},
  \bibinfo {author} {\bibfnamefont {Elliott~H.}\ \bibnamefont {Lieb}}, \ and\
  \bibinfo {author} {\bibfnamefont {Hal}\ \bibnamefont {Tasaki}},\ }\bibfield
  {title} {\enquote {\bibinfo {title} {Rigorous results on valence-bond ground
  states in antiferromagnets},}\ }\href {\doibase 10.1103/PhysRevLett.59.799}
  {\bibfield  {journal} {\bibinfo  {journal} {Phys. Rev. Lett.}\ }\textbf
  {\bibinfo {volume} {59}},\ \bibinfo {pages} {799--802} (\bibinfo {year}
  {1987})}\BibitemShut {NoStop}%
\bibitem [{\citenamefont {Else}\ \emph {et~al.}(2012)\citenamefont {Else},
  \citenamefont {Schwarz}, \citenamefont {Bartlett},\ and\ \citenamefont
  {Doherty}}]{Else:2012}%
  \BibitemOpen
  \bibfield  {author} {\bibinfo {author} {\bibfnamefont {Dominic~V.}\
  \bibnamefont {Else}}, \bibinfo {author} {\bibfnamefont {Ilai}\ \bibnamefont
  {Schwarz}}, \bibinfo {author} {\bibfnamefont {Stephen~D.}\ \bibnamefont
  {Bartlett}}, \ and\ \bibinfo {author} {\bibfnamefont {Andrew~C.}\
  \bibnamefont {Doherty}},\ }\bibfield  {title} {\enquote {\bibinfo {title}
  {Symmetry-protected phases for measurement-based quantum computation},}\
  }\href {\doibase 10.1103/PhysRevLett.108.240505} {\bibfield  {journal}
  {\bibinfo  {journal} {Phys. Rev. Lett.}\ }\textbf {\bibinfo {volume} {108}},\
  \bibinfo {pages} {240505} (\bibinfo {year} {2012})}\BibitemShut {NoStop}%
\bibitem [{\citenamefont {Moudgalya}\ \emph {et~al.}(2018)\citenamefont
  {Moudgalya}, \citenamefont {Regnault},\ and\ \citenamefont
  {Bernevig}}]{Moudgalya:2018}%
  \BibitemOpen
  \bibfield  {author} {\bibinfo {author} {\bibfnamefont {Sanjay}\ \bibnamefont
  {Moudgalya}}, \bibinfo {author} {\bibfnamefont {Nicolas}\ \bibnamefont
  {Regnault}}, \ and\ \bibinfo {author} {\bibfnamefont {B.~Andrei}\
  \bibnamefont {Bernevig}},\ }\bibfield  {title} {\enquote {\bibinfo {title}
  {Entanglement of exact excited states of affleck-kennedy-lieb-tasaki models:
  Exact results, many-body scars, and violation of the strong eigenstate
  thermalization hypothesis},}\ }\href {\doibase 10.1103/PhysRevB.98.235156}
  {\bibfield  {journal} {\bibinfo  {journal} {Phys. Rev. B}\ }\textbf {\bibinfo
  {volume} {98}},\ \bibinfo {pages} {235156} (\bibinfo {year}
  {2018})}\BibitemShut {NoStop}%
\bibitem [{\citenamefont {Chattopadhyay}\ \emph {et~al.}(2020)\citenamefont
  {Chattopadhyay}, \citenamefont {Pichler}, \citenamefont {Lukin},\ and\
  \citenamefont {Ho}}]{Chattopadhyay:2020}%
  \BibitemOpen
  \bibfield  {author} {\bibinfo {author} {\bibfnamefont {Sambuddha}\
  \bibnamefont {Chattopadhyay}}, \bibinfo {author} {\bibfnamefont {Hannes}\
  \bibnamefont {Pichler}}, \bibinfo {author} {\bibfnamefont {Mikhail~D.}\
  \bibnamefont {Lukin}}, \ and\ \bibinfo {author} {\bibfnamefont {Wen~Wei}\
  \bibnamefont {Ho}},\ }\bibfield  {title} {\enquote {\bibinfo {title} {Quantum
  many-body scars from virtual entangled pairs},}\ }\href {\doibase
  10.1103/PhysRevB.101.174308} {\bibfield  {journal} {\bibinfo  {journal}
  {Phys. Rev. B}\ }\textbf {\bibinfo {volume} {101}},\ \bibinfo {pages}
  {174308} (\bibinfo {year} {2020})}\BibitemShut {NoStop}%
\bibitem [{\citenamefont {Pollmann}\ \emph {et~al.}(2010)\citenamefont
  {Pollmann}, \citenamefont {Turner}, \citenamefont {Berg},\ and\ \citenamefont
  {Oshikawa}}]{Pollman:2010}%
  \BibitemOpen
  \bibfield  {author} {\bibinfo {author} {\bibfnamefont {Frank}\ \bibnamefont
  {Pollmann}}, \bibinfo {author} {\bibfnamefont {Ari~M.}\ \bibnamefont
  {Turner}}, \bibinfo {author} {\bibfnamefont {Erez}\ \bibnamefont {Berg}}, \
  and\ \bibinfo {author} {\bibfnamefont {Masaki}\ \bibnamefont {Oshikawa}},\
  }\bibfield  {title} {\enquote {\bibinfo {title} {Entanglement spectrum of a
  topological phase in one dimension},}\ }\href {\doibase
  10.1103/PhysRevB.81.064439} {\bibfield  {journal} {\bibinfo  {journal} {Phys.
  Rev. B}\ }\textbf {\bibinfo {volume} {81}},\ \bibinfo {pages} {064439}
  (\bibinfo {year} {2010})}\BibitemShut {NoStop}%
\bibitem [{\citenamefont {Georges}\ \emph {et~al.}(2001)\citenamefont
  {Georges}, \citenamefont {Borrás-Almenar}, \citenamefont {Coronado},
  \citenamefont {Curély},\ and\ \citenamefont {Drillon}}]{Renard_Ch2:2001}%
  \BibitemOpen
  \bibfield  {author} {\bibinfo {author} {\bibfnamefont {Roland}\ \bibnamefont
  {Georges}}, \bibinfo {author} {\bibfnamefont {Juan~J.}\ \bibnamefont
  {Borrás-Almenar}}, \bibinfo {author} {\bibfnamefont {Eugenio}\ \bibnamefont
  {Coronado}}, \bibinfo {author} {\bibfnamefont {Jacques}\ \bibnamefont
  {Curély}}, \ and\ \bibinfo {author} {\bibfnamefont {Marc}\ \bibnamefont
  {Drillon}},\ }\enquote {\bibinfo {title} {One-dimensional magnetism: An
  overview of the models},}\ in\ \href {\doibase
  https://doi.org/10.1002/3527600841.ch1} {\emph {\bibinfo {booktitle}
  {Magnetism: Molecules to Materials I}}}\ (\bibinfo  {publisher} {John Wiley
  \& Sons, Ltd},\ \bibinfo {year} {2001})\ Chap.~\bibinfo {chapter} {1}, pp.\
  \bibinfo {pages} {1--47}\BibitemShut {NoStop}%
\bibitem [{\citenamefont {Chen}\ \emph {et~al.}(2023)\citenamefont {Chen},
  \citenamefont {Shen}, \citenamefont {Lee},\ and\ \citenamefont
  {Yang}}]{Chen:2023}%
  \BibitemOpen
  \bibfield  {author} {\bibinfo {author} {\bibfnamefont {Tianqi}\ \bibnamefont
  {Chen}}, \bibinfo {author} {\bibfnamefont {Ruizhe}\ \bibnamefont {Shen}},
  \bibinfo {author} {\bibfnamefont {Ching~Hua}\ \bibnamefont {Lee}}, \ and\
  \bibinfo {author} {\bibfnamefont {Bo}~\bibnamefont {Yang}},\ }\bibfield
  {title} {\enquote {\bibinfo {title} {{High-fidelity realization of the AKLT
  state on a NISQ-era quantum processor}},}\ }\href {\doibase
  10.21468/SciPostPhys.15.4.170} {\bibfield  {journal} {\bibinfo  {journal}
  {SciPost Phys.}\ }\textbf {\bibinfo {volume} {15}},\ \bibinfo {pages} {170}
  (\bibinfo {year} {2023})}\BibitemShut {NoStop}%
\bibitem [{\citenamefont {Sompet}\ \emph {et~al.}(2022)\citenamefont {Sompet},
  \citenamefont {Hirthe}, \citenamefont {Bourgund}, \citenamefont {Chalopin},
  \citenamefont {Bibo}, \citenamefont {Koepsell}, \citenamefont {Bojović},
  \citenamefont {Verresen}, \citenamefont {Pollmann}, \citenamefont {Salomon},
  \citenamefont {Gross}, \citenamefont {Hilker},\ and\ \citenamefont
  {Bloch}}]{Sompet:2022}%
  \BibitemOpen
  \bibfield  {author} {\bibinfo {author} {\bibfnamefont {Pimonpan}\
  \bibnamefont {Sompet}}, \bibinfo {author} {\bibfnamefont {Sarah}\
  \bibnamefont {Hirthe}}, \bibinfo {author} {\bibfnamefont {Dominik}\
  \bibnamefont {Bourgund}}, \bibinfo {author} {\bibfnamefont {Thomas}\
  \bibnamefont {Chalopin}}, \bibinfo {author} {\bibfnamefont {Julian}\
  \bibnamefont {Bibo}}, \bibinfo {author} {\bibfnamefont {Joannis}\
  \bibnamefont {Koepsell}}, \bibinfo {author} {\bibfnamefont {Petar}\
  \bibnamefont {Bojović}}, \bibinfo {author} {\bibfnamefont {Ruben}\
  \bibnamefont {Verresen}}, \bibinfo {author} {\bibfnamefont {Frank}\
  \bibnamefont {Pollmann}}, \bibinfo {author} {\bibfnamefont {Guillaume}\
  \bibnamefont {Salomon}}, \bibinfo {author} {\bibfnamefont {Christian}\
  \bibnamefont {Gross}}, \bibinfo {author} {\bibfnamefont {Timon~A.}\
  \bibnamefont {Hilker}}, \ and\ \bibinfo {author} {\bibfnamefont {Immanuel}\
  \bibnamefont {Bloch}},\ }\bibfield  {title} {\enquote {\bibinfo {title}
  {Realizing the symmetry-protected haldane phase in fermi–hubbard
  ladders},}\ }\href {\doibase 10.1038/s41586-022-04688-z} {\bibfield
  {journal} {\bibinfo  {journal} {Nature}\ }\textbf {\bibinfo {volume} {606}},\
  \bibinfo {pages} {484–488} (\bibinfo {year} {2022})}\BibitemShut {NoStop}%
\bibitem [{\citenamefont {Cohen}\ and\ \citenamefont
  {Retzker}(2014)}]{Cohen:2014}%
  \BibitemOpen
  \bibfield  {author} {\bibinfo {author} {\bibfnamefont {I.}~\bibnamefont
  {Cohen}}\ and\ \bibinfo {author} {\bibfnamefont {A.}~\bibnamefont
  {Retzker}},\ }\bibfield  {title} {\enquote {\bibinfo {title} {Proposal for
  verification of the haldane phase using trapped ions},}\ }\href {\doibase
  10.1103/PhysRevLett.112.040503} {\bibfield  {journal} {\bibinfo  {journal}
  {Phys. Rev. Lett.}\ }\textbf {\bibinfo {volume} {112}},\ \bibinfo {pages}
  {040503} (\bibinfo {year} {2014})}\BibitemShut {NoStop}%
\bibitem [{\citenamefont {Cohen}\ \emph {et~al.}(2015)\citenamefont {Cohen},
  \citenamefont {Richerme}, \citenamefont {Gong}, \citenamefont {Monroe},\ and\
  \citenamefont {Retzker}}]{Cohen:2015}%
  \BibitemOpen
  \bibfield  {author} {\bibinfo {author} {\bibfnamefont {I.}~\bibnamefont
  {Cohen}}, \bibinfo {author} {\bibfnamefont {P.}~\bibnamefont {Richerme}},
  \bibinfo {author} {\bibfnamefont {Z.-X.}\ \bibnamefont {Gong}}, \bibinfo
  {author} {\bibfnamefont {C.}~\bibnamefont {Monroe}}, \ and\ \bibinfo {author}
  {\bibfnamefont {A.}~\bibnamefont {Retzker}},\ }\bibfield  {title} {\enquote
  {\bibinfo {title} {Simulating the haldane phase in trapped-ion spins using
  optical fields},}\ }\href {\doibase 10.1103/PhysRevA.92.012334} {\bibfield
  {journal} {\bibinfo  {journal} {Phys. Rev. A}\ }\textbf {\bibinfo {volume}
  {92}},\ \bibinfo {pages} {012334} (\bibinfo {year} {2015})}\BibitemShut
  {NoStop}%
\bibitem [{\citenamefont {Senko}\ \emph {et~al.}(2015)\citenamefont {Senko},
  \citenamefont {Richerme}, \citenamefont {Smith}, \citenamefont {Lee},
  \citenamefont {Cohen}, \citenamefont {Retzker},\ and\ \citenamefont
  {Monroe}}]{Senko:2015}%
  \BibitemOpen
  \bibfield  {author} {\bibinfo {author} {\bibfnamefont {C.}~\bibnamefont
  {Senko}}, \bibinfo {author} {\bibfnamefont {P.}~\bibnamefont {Richerme}},
  \bibinfo {author} {\bibfnamefont {J.}~\bibnamefont {Smith}}, \bibinfo
  {author} {\bibfnamefont {A.}~\bibnamefont {Lee}}, \bibinfo {author}
  {\bibfnamefont {I.}~\bibnamefont {Cohen}}, \bibinfo {author} {\bibfnamefont
  {A.}~\bibnamefont {Retzker}}, \ and\ \bibinfo {author} {\bibfnamefont
  {C.}~\bibnamefont {Monroe}},\ }\bibfield  {title} {\enquote {\bibinfo {title}
  {Realization of a quantum integer-spin chain with controllable
  interactions},}\ }\href {\doibase 10.1103/PhysRevX.5.021026} {\bibfield
  {journal} {\bibinfo  {journal} {Phys. Rev. X}\ }\textbf {\bibinfo {volume}
  {5}},\ \bibinfo {pages} {021026} (\bibinfo {year} {2015})}\BibitemShut
  {NoStop}%
\bibitem [{\citenamefont {Wang}\ \emph {et~al.}(2023)\citenamefont {Wang},
  \citenamefont {Snizhko}, \citenamefont {Romito}, \citenamefont {Gefen},\ and\
  \citenamefont {Murch}}]{Wang:2023}%
  \BibitemOpen
  \bibfield  {author} {\bibinfo {author} {\bibfnamefont {Yunzhao}\ \bibnamefont
  {Wang}}, \bibinfo {author} {\bibfnamefont {Kyrylo}\ \bibnamefont {Snizhko}},
  \bibinfo {author} {\bibfnamefont {Alessandro}\ \bibnamefont {Romito}},
  \bibinfo {author} {\bibfnamefont {Yuval}\ \bibnamefont {Gefen}}, \ and\
  \bibinfo {author} {\bibfnamefont {Kater}\ \bibnamefont {Murch}},\ }\bibfield
  {title} {\enquote {\bibinfo {title} {Dissipative preparation and
  stabilization of many-body quantum states in a superconducting qutrit
  array},}\ }\href {\doibase 10.1103/PhysRevA.108.013712} {\bibfield  {journal}
  {\bibinfo  {journal} {Phys. Rev. A}\ }\textbf {\bibinfo {volume} {108}},\
  \bibinfo {pages} {013712} (\bibinfo {year} {2023})}\BibitemShut {NoStop}%
\bibitem [{\citenamefont {Zhou}\ \emph {et~al.}(2021)\citenamefont {Zhou},
  \citenamefont {Choi},\ and\ \citenamefont {Lukin}}]{Zhou:2021}%
  \BibitemOpen
  \bibfield  {author} {\bibinfo {author} {\bibfnamefont {L.}~\bibnamefont
  {Zhou}}, \bibinfo {author} {\bibfnamefont {S.}~\bibnamefont {Choi}}, \ and\
  \bibinfo {author} {\bibfnamefont {M.~D.}\ \bibnamefont {Lukin}},\ }\bibfield
  {title} {\enquote {\bibinfo {title} {Symmetry-protected dissipative
  preparation of matrix product states},}\ }\href {\doibase
  10.1103/PhysRevA.104.032418} {\bibfield  {journal} {\bibinfo  {journal}
  {Phys. Rev. A}\ }\textbf {\bibinfo {volume} {104}},\ \bibinfo {pages}
  {032418} (\bibinfo {year} {2021})}\BibitemShut {NoStop}%
\bibitem [{\citenamefont {Kaltenbaek}\ \emph {et~al.}(2010)\citenamefont
  {Kaltenbaek}, \citenamefont {Lavoie}, \citenamefont {Zeng}, \citenamefont
  {Bartlett},\ and\ \citenamefont {Resch}}]{Kaltenbaek:2010}%
  \BibitemOpen
  \bibfield  {author} {\bibinfo {author} {\bibfnamefont {Rainer}\ \bibnamefont
  {Kaltenbaek}}, \bibinfo {author} {\bibfnamefont {Jonathan}\ \bibnamefont
  {Lavoie}}, \bibinfo {author} {\bibfnamefont {Bei}\ \bibnamefont {Zeng}},
  \bibinfo {author} {\bibfnamefont {Stephen~D.}\ \bibnamefont {Bartlett}}, \
  and\ \bibinfo {author} {\bibfnamefont {Kevin~J.}\ \bibnamefont {Resch}},\
  }\bibfield  {title} {\enquote {\bibinfo {title} {Optical one-way quantum
  computing with a simulated valence-bond solid},}\ }\href {\doibase
  10.1038/nphys1777} {\bibfield  {journal} {\bibinfo  {journal} {Nature
  Physics}\ }\textbf {\bibinfo {volume} {6}},\ \bibinfo {pages} {850--854}
  (\bibinfo {year} {2010})}\BibitemShut {NoStop}%
\bibitem [{\citenamefont {Smith}\ \emph {et~al.}(2023)\citenamefont {Smith},
  \citenamefont {Crane}, \citenamefont {Wiebe},\ and\ \citenamefont
  {Girvin}}]{Smith:2023}%
  \BibitemOpen
  \bibfield  {author} {\bibinfo {author} {\bibfnamefont {Kevin~C.}\
  \bibnamefont {Smith}}, \bibinfo {author} {\bibfnamefont {Eleanor}\
  \bibnamefont {Crane}}, \bibinfo {author} {\bibfnamefont {Nathan}\
  \bibnamefont {Wiebe}}, \ and\ \bibinfo {author} {\bibfnamefont {S.M.}\
  \bibnamefont {Girvin}},\ }\bibfield  {title} {\enquote {\bibinfo {title}
  {Deterministic constant-depth preparation of the aklt state on a quantum
  processor using fusion measurements},}\ }\href {\doibase
  10.1103/PRXQuantum.4.020315} {\bibfield  {journal} {\bibinfo  {journal} {PRX
  Quantum}\ }\textbf {\bibinfo {volume} {4}},\ \bibinfo {pages} {020315}
  (\bibinfo {year} {2023})}\BibitemShut {NoStop}%
\bibitem [{\citenamefont {Lake}\ \emph {et~al.}(2009)\citenamefont {Lake},
  \citenamefont {Tsvelik}, \citenamefont {Notbohm}, \citenamefont
  {Alan~Tennant}, \citenamefont {Perring}, \citenamefont {Reehuis},
  \citenamefont {Sekar}, \citenamefont {Krabbes},\ and\ \citenamefont
  {B\"{u}chner}}]{Lake:2009}%
  \BibitemOpen
  \bibfield  {author} {\bibinfo {author} {\bibfnamefont {Bella}\ \bibnamefont
  {Lake}}, \bibinfo {author} {\bibfnamefont {Alexei~M.}\ \bibnamefont
  {Tsvelik}}, \bibinfo {author} {\bibfnamefont {Susanne}\ \bibnamefont
  {Notbohm}}, \bibinfo {author} {\bibfnamefont {D.}~\bibnamefont
  {Alan~Tennant}}, \bibinfo {author} {\bibfnamefont {Toby~G.}\ \bibnamefont
  {Perring}}, \bibinfo {author} {\bibfnamefont {Manfred}\ \bibnamefont
  {Reehuis}}, \bibinfo {author} {\bibfnamefont {Chinnathambi}\ \bibnamefont
  {Sekar}}, \bibinfo {author} {\bibfnamefont {Gernot}\ \bibnamefont {Krabbes}},
  \ and\ \bibinfo {author} {\bibfnamefont {Bernd}\ \bibnamefont
  {B\"{u}chner}},\ }\bibfield  {title} {\enquote {\bibinfo {title} {Confinement
  of fractional quantum number particles in a condensed-matter system},}\
  }\href {\doibase 10.1038/nphys1462} {\bibfield  {journal} {\bibinfo
  {journal} {Nature Physics}\ }\textbf {\bibinfo {volume} {6}},\ \bibinfo
  {pages} {50–55} (\bibinfo {year} {2009})}\BibitemShut {NoStop}%
\bibitem [{\citenamefont {Ringbauer}\ \emph {et~al.}(2022)\citenamefont
  {Ringbauer}, \citenamefont {Meth}, \citenamefont {Postler}, \citenamefont
  {Stricker}, \citenamefont {Blatt}, \citenamefont {Schindler},\ and\
  \citenamefont {Monz}}]{Ringbauer:2022}%
  \BibitemOpen
  \bibfield  {author} {\bibinfo {author} {\bibfnamefont {Martin}\ \bibnamefont
  {Ringbauer}}, \bibinfo {author} {\bibfnamefont {Michael}\ \bibnamefont
  {Meth}}, \bibinfo {author} {\bibfnamefont {Lukas}\ \bibnamefont {Postler}},
  \bibinfo {author} {\bibfnamefont {Roman}\ \bibnamefont {Stricker}}, \bibinfo
  {author} {\bibfnamefont {Rainer}\ \bibnamefont {Blatt}}, \bibinfo {author}
  {\bibfnamefont {Philipp}\ \bibnamefont {Schindler}}, \ and\ \bibinfo {author}
  {\bibfnamefont {Thomas}\ \bibnamefont {Monz}},\ }\bibfield  {title} {\enquote
  {\bibinfo {title} {A universal qudit quantum processor with trapped ions},}\
  }\href {\doibase 10.1038/s41567-022-01658-0} {\bibfield  {journal} {\bibinfo
  {journal} {Nature Physics}\ }\textbf {\bibinfo {volume} {18}},\ \bibinfo
  {pages} {1053–1057} (\bibinfo {year} {2022})}\BibitemShut {NoStop}%
\bibitem [{\citenamefont {Kristen}\ \emph {et~al.}(2020)\citenamefont
  {Kristen}, \citenamefont {Schneider}, \citenamefont {Stehli}, \citenamefont
  {Wolz}, \citenamefont {Danilin}, \citenamefont {Ku}, \citenamefont {Long},
  \citenamefont {Wu}, \citenamefont {Lake}, \citenamefont {Pappas},
  \citenamefont {Ustinov},\ and\ \citenamefont {Weides}}]{Kristen:2020}%
  \BibitemOpen
  \bibfield  {author} {\bibinfo {author} {\bibfnamefont {M.}~\bibnamefont
  {Kristen}}, \bibinfo {author} {\bibfnamefont {A.}~\bibnamefont {Schneider}},
  \bibinfo {author} {\bibfnamefont {A.}~\bibnamefont {Stehli}}, \bibinfo
  {author} {\bibfnamefont {T.}~\bibnamefont {Wolz}}, \bibinfo {author}
  {\bibfnamefont {S.}~\bibnamefont {Danilin}}, \bibinfo {author} {\bibfnamefont
  {H.~S.}\ \bibnamefont {Ku}}, \bibinfo {author} {\bibfnamefont
  {J.}~\bibnamefont {Long}}, \bibinfo {author} {\bibfnamefont {X.}~\bibnamefont
  {Wu}}, \bibinfo {author} {\bibfnamefont {R.}~\bibnamefont {Lake}}, \bibinfo
  {author} {\bibfnamefont {D.~P.}\ \bibnamefont {Pappas}}, \bibinfo {author}
  {\bibfnamefont {A.~V.}\ \bibnamefont {Ustinov}}, \ and\ \bibinfo {author}
  {\bibfnamefont {M.}~\bibnamefont {Weides}},\ }\bibfield  {title} {\enquote
  {\bibinfo {title} {Amplitude and frequency sensing of microwave fields with a
  superconducting transmon qudit},}\ }\href {\doibase
  10.1038/s41534-020-00287-w} {\bibfield  {journal} {\bibinfo  {journal} {npj
  Quantum Information}\ }\textbf {\bibinfo {volume} {6}} (\bibinfo {year}
  {2020}),\ 10.1038/s41534-020-00287-w}\BibitemShut {NoStop}%
\bibitem [{\citenamefont {Chi}\ \emph {et~al.}(2022)\citenamefont {Chi},
  \citenamefont {Huang}, \citenamefont {Zhang}, \citenamefont {Mao},
  \citenamefont {Zhou}, \citenamefont {Chen}, \citenamefont {Zhai},
  \citenamefont {Bao}, \citenamefont {Dai}, \citenamefont {Yuan}, \citenamefont
  {Zhang}, \citenamefont {Dai}, \citenamefont {Tang}, \citenamefont {Yang},
  \citenamefont {Li}, \citenamefont {Ding}, \citenamefont {Oxenl{\o}we},
  \citenamefont {Thompson}, \citenamefont {O'Brien}, \citenamefont {Li},
  \citenamefont {Gong},\ and\ \citenamefont {Wang}}]{Chi:2022}%
  \BibitemOpen
  \bibfield  {author} {\bibinfo {author} {\bibfnamefont {Yulin}\ \bibnamefont
  {Chi}}, \bibinfo {author} {\bibfnamefont {Jieshan}\ \bibnamefont {Huang}},
  \bibinfo {author} {\bibfnamefont {Zhanchuan}\ \bibnamefont {Zhang}}, \bibinfo
  {author} {\bibfnamefont {Jun}\ \bibnamefont {Mao}}, \bibinfo {author}
  {\bibfnamefont {Zinan}\ \bibnamefont {Zhou}}, \bibinfo {author}
  {\bibfnamefont {Xiaojiong}\ \bibnamefont {Chen}}, \bibinfo {author}
  {\bibfnamefont {Chonghao}\ \bibnamefont {Zhai}}, \bibinfo {author}
  {\bibfnamefont {Jueming}\ \bibnamefont {Bao}}, \bibinfo {author}
  {\bibfnamefont {Tianxiang}\ \bibnamefont {Dai}}, \bibinfo {author}
  {\bibfnamefont {Huihong}\ \bibnamefont {Yuan}}, \bibinfo {author}
  {\bibfnamefont {Ming}\ \bibnamefont {Zhang}}, \bibinfo {author}
  {\bibfnamefont {Daoxin}\ \bibnamefont {Dai}}, \bibinfo {author}
  {\bibfnamefont {Bo}~\bibnamefont {Tang}}, \bibinfo {author} {\bibfnamefont
  {Yan}\ \bibnamefont {Yang}}, \bibinfo {author} {\bibfnamefont {Zhihua}\
  \bibnamefont {Li}}, \bibinfo {author} {\bibfnamefont {Yunhong}\ \bibnamefont
  {Ding}}, \bibinfo {author} {\bibfnamefont {Leif~K.}\ \bibnamefont
  {Oxenl{\o}we}}, \bibinfo {author} {\bibfnamefont {Mark~G.}\ \bibnamefont
  {Thompson}}, \bibinfo {author} {\bibfnamefont {Jeremy~L.}\ \bibnamefont
  {O'Brien}}, \bibinfo {author} {\bibfnamefont {Yan}\ \bibnamefont {Li}},
  \bibinfo {author} {\bibfnamefont {Qihuang}\ \bibnamefont {Gong}}, \ and\
  \bibinfo {author} {\bibfnamefont {Jianwei}\ \bibnamefont {Wang}},\ }\bibfield
   {title} {\enquote {\bibinfo {title} {A programmable qudit-based quantum
  processor},}\ }\href@noop {} {\bibfield  {journal} {\bibinfo  {journal} {Nat.
  Commun.}\ }\textbf {\bibinfo {volume} {13}},\ \bibinfo {pages} {1166}
  (\bibinfo {year} {2022})}\BibitemShut {NoStop}%
\bibitem [{\citenamefont {Cirac}\ \emph {et~al.}(2021)\citenamefont {Cirac},
  \citenamefont {P\'erez-Garc\'{\i}a}, \citenamefont {Schuch},\ and\
  \citenamefont {Verstraete}}]{Cirac:2021}%
  \BibitemOpen
  \bibfield  {author} {\bibinfo {author} {\bibfnamefont {J.~Ignacio}\
  \bibnamefont {Cirac}}, \bibinfo {author} {\bibfnamefont {David}\ \bibnamefont
  {P\'erez-Garc\'{\i}a}}, \bibinfo {author} {\bibfnamefont {Norbert}\
  \bibnamefont {Schuch}}, \ and\ \bibinfo {author} {\bibfnamefont {Frank}\
  \bibnamefont {Verstraete}},\ }\bibfield  {title} {\enquote {\bibinfo {title}
  {Matrix product states and projected entangled pair states: Concepts,
  symmetries, theorems},}\ }\href {\doibase 10.1103/RevModPhys.93.045003}
  {\bibfield  {journal} {\bibinfo  {journal} {Rev. Mod. Phys.}\ }\textbf
  {\bibinfo {volume} {93}},\ \bibinfo {pages} {045003} (\bibinfo {year}
  {2021})}\BibitemShut {NoStop}%
\bibitem [{\citenamefont {Schuch}\ \emph {et~al.}(2011)\citenamefont {Schuch},
  \citenamefont {P\'erez-Garc\'{\i}a},\ and\ \citenamefont
  {Cirac}}]{Schuch:2011}%
  \BibitemOpen
  \bibfield  {author} {\bibinfo {author} {\bibfnamefont {Norbert}\ \bibnamefont
  {Schuch}}, \bibinfo {author} {\bibfnamefont {David}\ \bibnamefont
  {P\'erez-Garc\'{\i}a}}, \ and\ \bibinfo {author} {\bibfnamefont {Ignacio}\
  \bibnamefont {Cirac}},\ }\bibfield  {title} {\enquote {\bibinfo {title}
  {Classifying quantum phases using matrix product states and projected
  entangled pair states},}\ }\href {\doibase 10.1103/PhysRevB.84.165139}
  {\bibfield  {journal} {\bibinfo  {journal} {Phys. Rev. B}\ }\textbf {\bibinfo
  {volume} {84}},\ \bibinfo {pages} {165139} (\bibinfo {year}
  {2011})}\BibitemShut {NoStop}%
\bibitem [{\citenamefont {Sch\"on}\ \emph {et~al.}(2005)\citenamefont
  {Sch\"on}, \citenamefont {Solano}, \citenamefont {Verstraete}, \citenamefont
  {Cirac},\ and\ \citenamefont {Wolf}}]{Schon:2005}%
  \BibitemOpen
  \bibfield  {author} {\bibinfo {author} {\bibfnamefont {C.}~\bibnamefont
  {Sch\"on}}, \bibinfo {author} {\bibfnamefont {E.}~\bibnamefont {Solano}},
  \bibinfo {author} {\bibfnamefont {F.}~\bibnamefont {Verstraete}}, \bibinfo
  {author} {\bibfnamefont {J.~I.}\ \bibnamefont {Cirac}}, \ and\ \bibinfo
  {author} {\bibfnamefont {M.~M.}\ \bibnamefont {Wolf}},\ }\bibfield  {title}
  {\enquote {\bibinfo {title} {Sequential generation of entangled multiqubit
  states},}\ }\href {\doibase 10.1103/PhysRevLett.95.110503} {\bibfield
  {journal} {\bibinfo  {journal} {Phys. Rev. Lett.}\ }\textbf {\bibinfo
  {volume} {95}},\ \bibinfo {pages} {110503} (\bibinfo {year}
  {2005})}\BibitemShut {NoStop}%
\bibitem [{\citenamefont {Sch\"on}\ \emph {et~al.}(2007)\citenamefont
  {Sch\"on}, \citenamefont {Hammerer}, \citenamefont {Wolf}, \citenamefont
  {Cirac},\ and\ \citenamefont {Solano}}]{Schon:2007}%
  \BibitemOpen
  \bibfield  {author} {\bibinfo {author} {\bibfnamefont {C.}~\bibnamefont
  {Sch\"on}}, \bibinfo {author} {\bibfnamefont {K.}~\bibnamefont {Hammerer}},
  \bibinfo {author} {\bibfnamefont {M.~M.}\ \bibnamefont {Wolf}}, \bibinfo
  {author} {\bibfnamefont {J.~I.}\ \bibnamefont {Cirac}}, \ and\ \bibinfo
  {author} {\bibfnamefont {E.}~\bibnamefont {Solano}},\ }\bibfield  {title}
  {\enquote {\bibinfo {title} {Sequential generation of matrix-product states
  in cavity qed},}\ }\href {\doibase 10.1103/PhysRevA.75.032311} {\bibfield
  {journal} {\bibinfo  {journal} {Phys. Rev. A}\ }\textbf {\bibinfo {volume}
  {75}},\ \bibinfo {pages} {032311} (\bibinfo {year} {2007})}\BibitemShut
  {NoStop}%
\bibitem [{\citenamefont {Younis}\ \emph {et~al.}(2021)\citenamefont {Younis},
  \citenamefont {Iancu}, \citenamefont {Lavrijsen}, \citenamefont {Davis},
  \citenamefont {Smith},\ and\ \citenamefont {USDOE}}]{BQSKit}%
  \BibitemOpen
  \bibfield  {author} {\bibinfo {author} {\bibfnamefont {Ed}~\bibnamefont
  {Younis}}, \bibinfo {author} {\bibfnamefont {Costin~C}\ \bibnamefont
  {Iancu}}, \bibinfo {author} {\bibfnamefont {Wim}\ \bibnamefont {Lavrijsen}},
  \bibinfo {author} {\bibfnamefont {Marc}\ \bibnamefont {Davis}}, \bibinfo
  {author} {\bibfnamefont {Ethan}\ \bibnamefont {Smith}}, \ and\ \bibinfo
  {author} {\bibnamefont {USDOE}},\ }\href {\doibase 10.11578/dc.20210603.2}
  {\enquote {\bibinfo {title} {Berkeley quantum synthesis toolkit (bqskit)
  v1},}\ } (\bibinfo {year} {2021})\BibitemShut {NoStop}%
\bibitem [{\citenamefont {Kennedy}(1990)}]{Kennedy:1990}%
  \BibitemOpen
  \bibfield  {author} {\bibinfo {author} {\bibfnamefont {T}~\bibnamefont
  {Kennedy}},\ }\bibfield  {title} {\enquote {\bibinfo {title} {Exact
  diagonalisations of open spin-1 chains},}\ }\href {\doibase
  10.1088/0953-8984/2/26/010} {\bibfield  {journal} {\bibinfo  {journal}
  {Journal of Physics: Condensed Matter}\ }\textbf {\bibinfo {volume} {2}},\
  \bibinfo {pages} {5737–5745} (\bibinfo {year} {1990})}\BibitemShut
  {NoStop}%
\bibitem [{\citenamefont {P\'erez-Garc\'{\i}a}\ \emph
  {et~al.}(2008)\citenamefont {P\'erez-Garc\'{\i}a}, \citenamefont {Wolf},
  \citenamefont {Sanz}, \citenamefont {Verstraete},\ and\ \citenamefont
  {Cirac}}]{Perez:2008}%
  \BibitemOpen
  \bibfield  {author} {\bibinfo {author} {\bibfnamefont {D.}~\bibnamefont
  {P\'erez-Garc\'{\i}a}}, \bibinfo {author} {\bibfnamefont {M.~M.}\
  \bibnamefont {Wolf}}, \bibinfo {author} {\bibfnamefont {M.}~\bibnamefont
  {Sanz}}, \bibinfo {author} {\bibfnamefont {F.}~\bibnamefont {Verstraete}}, \
  and\ \bibinfo {author} {\bibfnamefont {J.~I.}\ \bibnamefont {Cirac}},\
  }\bibfield  {title} {\enquote {\bibinfo {title} {String order and symmetries
  in quantum spin lattices},}\ }\href {\doibase 10.1103/PhysRevLett.100.167202}
  {\bibfield  {journal} {\bibinfo  {journal} {Phys. Rev. Lett.}\ }\textbf
  {\bibinfo {volume} {100}},\ \bibinfo {pages} {167202} (\bibinfo {year}
  {2008})}\BibitemShut {NoStop}%
\bibitem [{\citenamefont {Nijs}\ and\ \citenamefont
  {Rommelse}(1989)}]{Nijs:1989}%
  \BibitemOpen
  \bibfield  {author} {\bibinfo {author} {\bibfnamefont {Marcel}\ \bibnamefont
  {Nijs}}\ and\ \bibinfo {author} {\bibfnamefont {Koos}\ \bibnamefont
  {Rommelse}},\ }\bibfield  {title} {\enquote {\bibinfo {title} {Preroughening
  transitions in crystal surfaces and valence-bond phases in quantum spin
  chains},}\ }\href {\doibase 10.1103/PhysRevB.40.4709} {\bibfield  {journal}
  {\bibinfo  {journal} {Phys. Rev. B}\ }\textbf {\bibinfo {volume} {40}},\
  \bibinfo {pages} {4709} (\bibinfo {year} {1989})}\BibitemShut {NoStop}%
\bibitem [{\citenamefont {Stricker}\ \emph {et~al.}(2022)\citenamefont
  {Stricker}, \citenamefont {Meth}, \citenamefont {Postler}, \citenamefont
  {Edmunds}, \citenamefont {Ferrie}, \citenamefont {Blatt}, \citenamefont
  {Schindler}, \citenamefont {Monz}, \citenamefont {Kueng},\ and\ \citenamefont
  {Ringbauer}}]{Stricker:2022}%
  \BibitemOpen
  \bibfield  {author} {\bibinfo {author} {\bibfnamefont {Roman}\ \bibnamefont
  {Stricker}}, \bibinfo {author} {\bibfnamefont {Michael}\ \bibnamefont
  {Meth}}, \bibinfo {author} {\bibfnamefont {Lukas}\ \bibnamefont {Postler}},
  \bibinfo {author} {\bibfnamefont {Claire}\ \bibnamefont {Edmunds}}, \bibinfo
  {author} {\bibfnamefont {Chris}\ \bibnamefont {Ferrie}}, \bibinfo {author}
  {\bibfnamefont {Rainer}\ \bibnamefont {Blatt}}, \bibinfo {author}
  {\bibfnamefont {Philipp}\ \bibnamefont {Schindler}}, \bibinfo {author}
  {\bibfnamefont {Thomas}\ \bibnamefont {Monz}}, \bibinfo {author}
  {\bibfnamefont {Richard}\ \bibnamefont {Kueng}}, \ and\ \bibinfo {author}
  {\bibfnamefont {Martin}\ \bibnamefont {Ringbauer}},\ }\bibfield  {title}
  {\enquote {\bibinfo {title} {Experimental single-setting quantum state
  tomography},}\ }\href {\doibase 10.1103/PRXQuantum.3.040310} {\bibfield
  {journal} {\bibinfo  {journal} {PRX Quantum}\ }\textbf {\bibinfo {volume}
  {3}},\ \bibinfo {pages} {040310} (\bibinfo {year} {2022})}\BibitemShut
  {NoStop}%
\bibitem [{\citenamefont {Hastings}\ and\ \citenamefont
  {Koma}(2006)}]{Hastings:2006}%
  \BibitemOpen
  \bibfield  {author} {\bibinfo {author} {\bibfnamefont {Matthew~B.}\
  \bibnamefont {Hastings}}\ and\ \bibinfo {author} {\bibfnamefont {Tohru}\
  \bibnamefont {Koma}},\ }\bibfield  {title} {\enquote {\bibinfo {title}
  {Spectral gap and exponential decay of correlations},}\ }\href {\doibase
  10.1007/s00220-006-0030-4} {\bibfield  {journal} {\bibinfo  {journal}
  {Communications in Mathematical Physics}\ }\textbf {\bibinfo {volume}
  {265}},\ \bibinfo {pages} {781–804} (\bibinfo {year} {2006})}\BibitemShut
  {NoStop}%
\bibitem [{\citenamefont {Sadoune}\ and\ \citenamefont
  {et~al.}(2024)}]{Sadoune:2024}%
  \BibitemOpen
  \bibfield  {author} {\bibinfo {author} {\bibfnamefont {N.}~\bibnamefont
  {Sadoune}}\ and\ \bibinfo {author} {\bibnamefont {et~al.}},\ }\href@noop {}
  {\enquote {\bibinfo {title} {[manuscript in preparation]},}\ } (\bibinfo
  {year} {2024})\BibitemShut {NoStop}%
\bibitem [{\citenamefont {Raussendorf}\ and\ \citenamefont
  {Briegel}(2001)}]{Raussendorf:2001}%
  \BibitemOpen
  \bibfield  {author} {\bibinfo {author} {\bibfnamefont {Robert}\ \bibnamefont
  {Raussendorf}}\ and\ \bibinfo {author} {\bibfnamefont {Hans~J.}\ \bibnamefont
  {Briegel}},\ }\bibfield  {title} {\enquote {\bibinfo {title} {A one-way
  quantum computer},}\ }\href {\doibase 10.1103/PhysRevLett.86.5188} {\bibfield
   {journal} {\bibinfo  {journal} {Phys. Rev. Lett.}\ }\textbf {\bibinfo
  {volume} {86}},\ \bibinfo {pages} {5188--5191} (\bibinfo {year}
  {2001})}\BibitemShut {NoStop}%
\bibitem [{\citenamefont {Raussendorf}\ \emph {et~al.}(2003)\citenamefont
  {Raussendorf}, \citenamefont {Browne},\ and\ \citenamefont
  {Briegel}}]{Raussendorf:2003}%
  \BibitemOpen
  \bibfield  {author} {\bibinfo {author} {\bibfnamefont {Robert}\ \bibnamefont
  {Raussendorf}}, \bibinfo {author} {\bibfnamefont {Daniel~E.}\ \bibnamefont
  {Browne}}, \ and\ \bibinfo {author} {\bibfnamefont {Hans~J.}\ \bibnamefont
  {Briegel}},\ }\bibfield  {title} {\enquote {\bibinfo {title}
  {Measurement-based quantum computation on cluster states},}\ }\href {\doibase
  10.1103/PhysRevA.68.022312} {\bibfield  {journal} {\bibinfo  {journal} {Phys.
  Rev. A}\ }\textbf {\bibinfo {volume} {68}},\ \bibinfo {pages} {022312}
  (\bibinfo {year} {2003})}\BibitemShut {NoStop}%
\bibitem [{\citenamefont {Ringbauer}\ \emph {et~al.}(2023)\citenamefont
  {Ringbauer}, \citenamefont {Hinsche}, \citenamefont {Feldker}, \citenamefont
  {Faehrmann}, \citenamefont {Bermejo-Vega}, \citenamefont {Edmunds},
  \citenamefont {Postler}, \citenamefont {R.}, \citenamefont {Marciniak},
  \citenamefont {Meth}, \citenamefont {Pogorelov}, \citenamefont {Blatt},
  \citenamefont {Schindler}, \citenamefont {Eisert}, \citenamefont {Monz},\
  and\ \citenamefont {Hangleiter}}]{Ringbauer:2023}%
  \BibitemOpen
  \bibfield  {author} {\bibinfo {author} {\bibfnamefont {M.}~\bibnamefont
  {Ringbauer}}, \bibinfo {author} {\bibfnamefont {M.}~\bibnamefont {Hinsche}},
  \bibinfo {author} {\bibfnamefont {T.}~\bibnamefont {Feldker}}, \bibinfo
  {author} {\bibfnamefont {P.~K.}\ \bibnamefont {Faehrmann}}, \bibinfo {author}
  {\bibfnamefont {J.}~\bibnamefont {Bermejo-Vega}}, \bibinfo {author}
  {\bibfnamefont {C.}~\bibnamefont {Edmunds}}, \bibinfo {author} {\bibfnamefont
  {L.}~\bibnamefont {Postler}}, \bibinfo {author} {\bibfnamefont {Stricker}\
  \bibnamefont {R.}}, \bibinfo {author} {\bibfnamefont {C.~D.}\ \bibnamefont
  {Marciniak}}, \bibinfo {author} {\bibfnamefont {M.}~\bibnamefont {Meth}},
  \bibinfo {author} {\bibfnamefont {I.}~\bibnamefont {Pogorelov}}, \bibinfo
  {author} {\bibfnamefont {R.}~\bibnamefont {Blatt}}, \bibinfo {author}
  {\bibfnamefont {P.}~\bibnamefont {Schindler}}, \bibinfo {author}
  {\bibfnamefont {J.}~\bibnamefont {Eisert}}, \bibinfo {author} {\bibfnamefont
  {T.}~\bibnamefont {Monz}}, \ and\ \bibinfo {author} {\bibfnamefont
  {D.}~\bibnamefont {Hangleiter}},\ }\href@noop {} {\enquote {\bibinfo {title}
  {Verifiable measurement-based quantum random sampling with trapped ions},}\ }
  (\bibinfo {year} {2023}),\ \Eprint {http://arxiv.org/abs/2307.14424}
  {arXiv:2307.14424} \BibitemShut {NoStop}%
\bibitem [{\citenamefont {Wei}\ \emph {et~al.}(2022)\citenamefont {Wei},
  \citenamefont {Malz},\ and\ \citenamefont {Cirac}}]{Wei:2022}%
  \BibitemOpen
  \bibfield  {author} {\bibinfo {author} {\bibfnamefont {Zhi-Yuan}\
  \bibnamefont {Wei}}, \bibinfo {author} {\bibfnamefont {Daniel}\ \bibnamefont
  {Malz}}, \ and\ \bibinfo {author} {\bibfnamefont {J.~Ignacio}\ \bibnamefont
  {Cirac}},\ }\bibfield  {title} {\enquote {\bibinfo {title} {Sequential
  generation of projected entangled-pair states},}\ }\href {\doibase
  10.1103/PhysRevLett.128.010607} {\bibfield  {journal} {\bibinfo  {journal}
  {Phys. Rev. Lett.}\ }\textbf {\bibinfo {volume} {128}},\ \bibinfo {pages}
  {010607} (\bibinfo {year} {2022})}\BibitemShut {NoStop}%
\end{thebibliography}%
\end{document}